\documentclass[%
 reprint,
 superscriptaddress,
 amsmath,amssymb,
 aps,
]{revtex4-2}

\usepackage{graphicx}
\usepackage{dcolumn}
\usepackage{bm}
\usepackage{color}
\usepackage{hyperref}

\begin{document}

\title{Damping Separation of Finite Open Systems in Gravity-Related Experiments in the Free Molecular Flow Regime}

\author{Hou-Qiang Teng}
\affiliation{Lanzhou Center for Theoretical Physics, Key Laboratory of Theoretical Physics of Gansu Province, and Key Laboratory of Quantum Theory and Applications of MoE, Lanzhou University, Lanzhou, Gansu 730000, China}
\author{Jia-Qi Dong}
\email{dongjq@lzu.edu.cn}
\affiliation{Lanzhou Center for Theoretical Physics, Key Laboratory of Theoretical Physics of Gansu Province, and Key Laboratory of Quantum Theory and Applications of MoE, Lanzhou University, Lanzhou, Gansu 730000, China}
\author{Yisen Wang}
\affiliation{Lanzhou Center for Theoretical Physics, Key Laboratory of Theoretical Physics of Gansu Province, and Key Laboratory of Quantum Theory and Applications of MoE, Lanzhou University, Lanzhou, Gansu 730000, China}
\author{Liang Huang}
\email{huangl@lzu.edu.cn}
\affiliation{Lanzhou Center for Theoretical Physics, Key Laboratory of Theoretical Physics of Gansu Province, and Key Laboratory of Quantum Theory and Applications of MoE, Lanzhou University, Lanzhou, Gansu 730000, China}
\author{Peng Xu}
\affiliation{Lanzhou Center for Theoretical Physics, Key Laboratory of Theoretical Physics of Gansu Province, and Key Laboratory of Quantum Theory and Applications of MoE, Lanzhou University, Lanzhou, Gansu 730000, China}
\affiliation{Center for Gravitational Wave Experiment, National Microgravity Laboratory, Institute of Mechanics, Chinese Academy of Sciences, Beijing 100190, China}
\affiliation{Hangzhou Institute for Advanced Study, University of Chinese Academy of Sciences, Hangzhou 310124, China}

\date{\today}

\begin{abstract}
The residual gas damping of the test mass (TM) in the free molecular flow regime is studied in the finite open systems for high-precision gravity-related experiments. Through strict derivation, we separate the damping coefficients for two finite open systems, i.e., the bi-plate system and the sensor core system, into base damping and diffusion damping. This elucidates the relationship between the free damping in the infinite gas volume and the proximity damping in the constrained volume, unifies them into one microscopic picture, and allows us to point out three pathways of energy dissipation in the bi-plate gap. We also provide the conditions that need to be met to achieve this separation. In applications, for space gravitational wave detection, our results for the residual gas damping coefficient for the 4TM torsion balance experiment is the closest one to the experimental and simulation data compared to previous models.  For the LISA mission, our estimation for residual gas acceleration noise at the sensitive axis is consistent with the simulation result, within about $5\%$ difference. In addition, in the test of the gravitational inverse-square law, our results suggest that the constraint on the distance between TM and the conducting membrane can be reduced by about $28\%$. 
\end{abstract}

\maketitle

\section{\label{sec:level1}Introduction}

Einstein’s general relativity, which still stands as the most successful theory of gravitation, has a far-reaching influence on the vision of nature as geometry. With the establishment of the Dicke framework~\cite{dicke1964experimental} based on Einstein’s equivalence principle (EP) in the 1960s, experimental gravity or gravitational experiments can be roughly divided into two classes, including the tests of foundations of gravitational theory such as EP and the precision measurements of spacetime curvature effects of the so-called metric theories of gravity~\cite{will2014confrontation,will2018theory}. Both classes of gravitational experiments heavily rely on the establishments of high-precision inertial references and the measurements of the relative free-falling motions or geodesic deviations between them, and these include, for example, the MICROSCOPE mission~\cite{TOUBOUL2002433,PhysRevLett.129.121102,PhysRevLett.119.231101} for the test of the weak EP, ground-based or space-borne gravitational wave (GW) antennas like the LIGO-VIRGO collaboration~\cite{PhysRevLett.116.061102,PhysRevD.102.062003,PhysRevX.13.041021}, the LISA~\cite{PhysRevLett.116.231101,PhysRevLett.120.061101} and LISA-like missions~\cite{Luo_2016,10.1093/nsr/nwx116}, satellite gravity recovery missions such as GRACE/GRACE-FO~\cite{naeimi2017global,10.2514/1.A34326} and GOCE~\cite{touboul2012champ}, and ground-based experiments with high-precision torsion balances~\cite{PhysRevLett.98.131104,PhysRevLett.108.081101, PhysRevLett.116.131101, PhysRevLett.124.051301, PhysRevLett.117.071102, PhysRevLett.122.011102, PhysRevLett.126.211101,li2018measurements,RevModPhys.93.025010,rosi2014precision} et al. In present days, isolated test masses (TMs) coupled with high-precision readout systems, such as capacity sensors~\cite{touboul1999accelerometers}, laser interferometers~\cite{doi:10.1126/science.256.5055.325}, or SQUIDs~\cite{doi:10.1142/S0218271819400017}, are the main technical implementation methods for high-precision inertial reference systems. The TM, as the key unit, is generally suspended (electrically or magnetically in space, or through a wire for ground-based experiments) inside a very stable environment with extremely weak couplings to external disturbances, which ensures the free-falling state of certain degrees of freedom of the TM in the gravitational field. While, the unavoidable fluctuations of the physical fields coupled to the TM, especially the random collisions from residual gas molecules, will limit the acceleration noise of TM and therefore the precisions or sensitivities of the inertial reference~\cite{BonnyLSchumaker_2003}. Therefore, the accurate assessments of numerous environmental noises become a crucial task in the high-precision gravity-related experiments. 

Among the various stray forces~\cite{BonnyLSchumaker_2003}, those associated with residual gases include damping force noise~\cite{PhysRevLett.103.140601,PhysRevD.84.063007}, radiometer effect, and outgassing effect~\cite{PhysRevD.76.102003,DatongXue_2011}. Damping force noise refers to the Brownian motion of TMs due to gas collisions, which typically becomes one of the limits in high-precision measurements~\cite{PhysRevLett.103.140601, PhysRevD.84.063007, PhysRevD.81.123008, CAVALLERI20103365, Ke_2020, Ke_2022, Mao_2023, PhysRevApplied.19.044005}. The damping force noise can be obtained from the damping coefficient through the fluctuation-dissipation theorem~\cite{PhysRev.86.702,PhysRev.91.1505,R_Kubo_1966,PhysRevD.42.2437,PhysRevLett.103.140601}
\begin{equation}
	S_{f}(\omega)=4k_{\mathrm{B}}T\mathrm{Re}\left(-\frac{\partial F(\omega)}{\partial v(\omega)}\right)=4k_{\mathrm{B}} T\mathrm{Re}[Z(\omega)],
	\label{eq:fluctuationdamping}
\end{equation}
where $S_{f}(\omega)$ is the power spectral density of the fluctuation force on the TM, $k_{\mathrm{B}}$ is the Boltzmann constant, $T$ is the temperature, $F$ is the damping force, $v$ is the TM's velocity, $Z(\omega)$ is the mechanical impedance of the system. In high-precision gravity-related experiments, to reduce residual gas noise, the environment pressure is generally less than $10^{-4}\ \mathrm{Pa}$~\cite{DatongXue_2011,PhysRevLett.116.131101,PhysRevD.106.062001}, and the environment around the TM is in a free molecular flow regime, where the mean free path is much larger than the distance from the TM to the surrounding walls, and collisions between molecules rarely occur. The collisions between molecules and the surfaces are inelastic, following the Knudsen hypothesis~\cite{andp.19093330106,knudsen1967cosine,knudsen1911molekularstromung,knudsen1917vaporisation,knudsen1917vaporisation}. So the collisions can be treated as independent impulses, the resulting fluctuation force noise has a frequency independent spectrum~\cite{PhysRevD.84.063007}, so residual gas leads to an impedance $Z(\omega)=\beta$, with $\beta$ referred to here as the gas damping coefficient.

The residual gas damping coefficient is dependent on the specific environments surrounding TMs. In high-precision gravity-related experimental setups with TMs, two fundamental structures frequently appear, i.e., bi-plate and sensor core~\cite{Manuel_Rodrigues_2003,marque2008ultra}. The bi-plate structure comprises a movable TM and a fixed parallel plate. The sensor core is a cuboid nested structure within gravitational reference sensors (GRSs)/inertial sensors, with the TM in the middle and surrounded by an electrode housing. And they constitute the finite open systems. In GW detections, the damping caused by the motion of the TM is called proximity-enhanced damping~\cite{PhysRevD.84.063007, Ke_2022, Mao_2023}, while in other fields like MEMS/NEMS, it is known as squeeze-film damping~\cite{SUIJLEN2009171, acs.nanolett.1c02237}. In traditional squeeze-film damping studies, when the pressure is at one atmospheric pressure or less, the gas is treated as a continuous viscous fluid, and damping forces are calculated using the viscosity coefficient or effective viscosity coefficient~\cite{10.1117/12.395618,ANDREWS199379,ANDREWS1995103}. As the pressure decreases until collisions between molecules rarely occur, i.e. the free molecular flow regime, the environment around TM and the choice of surface boundary conditions and analysis methods significantly influence the damping calculations. Commonly employed methods include the free molecular model~\cite{721736,LI1999191,ZOOK199251,MinhangBao_2002,Lu_2018}, isothermal piston model~\cite{SUIJLEN2009171,PhysRevD.84.063007}, shot noise model~\cite{PhysRevD.84.063007}, and non-isothermal models~\cite{5398187,Ke_2022}.

In traditional analyses, the damping coefficient $\beta$ in the finite open systems consists of two parts~\cite{PhysRevD.84.063007,Ke_2020,SUIJLEN2009171,Mao_2023}, i.e., the free damping/kinetic damping $\beta_{\infty}$ within an infinite gas volume, and proximity damping $\Delta\beta$ within a constrained volume. And the total damping coefficient is $\beta=\beta_{\infty}+\Delta\beta$, corresponding to the force noise power spectral density, $S_{f}(\omega)=S_{f}^{\infty}+\Delta S_{f}(\omega)$. However, for this linear separation, there is no consistent understanding of the relationship between the two dampings. For example, 
Ref.~\cite{PhysRevD.84.063007} gives a simple explanation: there is a continuum of behavior between the free damping limit and proximity damping, which is confirmed by Monte Carlo simulations. Reference~\cite{SUIJLEN2009171} compares the magnitudes of two dampings, and kinetic damping is eventually ignored. And Ref.~\cite{Mao_2023} considers only the proximity damping on the front side, with shear free damping on the lateral side. Therefore, a quantitative and rigorous analysis for the separation of the damping and the corresponding condition is needed.

To address the above issues, some researchers have studied from the perspective of non-isothermal thermodynamics. Ref.~\cite{5398187} considers the impact of gas temperature increase during the compression process, introduces the thermal-squeeze film damping, and highlights the influence of molecular degrees of freedom and surface roughness on damping. However, their use of adiabatic conditions contradicts the Knudsen hypothesis. Subsequently, Ref.~\cite{Ke_2022} introduces isothermal conditions, and employs effective temperature to compensate for the excess internal energy in the non-equilibrium state compared to the equilibrium state. While effective temperature is a macroscopic approximation, and high-precision measurements demand a more microscopically accurate explanation for residual gas noise.

In this paper, we introduce finite open models, and precisely decompose the damping coefficient into base damping and diffusion damping for both bi-plate and sensor core finite open systems. This clarifies the relationship between the free damping and the proximity damping, and unifies them into one microscopic picture. In applications, the finite open models effectively correct the constraints on gap spacing in the test of the gravitational inverse-square law (ISL), and emerge as the theoretical model most aligned with the simulation and measurement results of the residual gas damping in GRSs for the LISA mission and other LISA-like space GW detection missions.

The rest of this paper is organized as follows: Section~\ref{sec:density} calculates the changes in molecular number density. The separation of damping coefficients is carried out for both bi-plate and sensor core finite open systems in Sec.~\ref{sec:seperation}. Section~\ref{sec:energy} gives separation and sources of damping from an energy perspective. The finite open models are applied to experiments in the test of the ISL and the GW detections in Sec.~\ref{sec:application}. Discussions and conclusions are provided in Sec.~\ref{sec:conclusion}. Appendices include detailed derivations and expressions for some physical quantities.

\section{\label{sec:density} Changes in molecular number density in the bi-plate finite open system}

The distinction between the bi-plate finite open system and the infinite volume lies in: as the TM moves, the molecular number density within the gap changes. When calculating the momentum exchange between molecules and the TM, the number density is required. This will lead to the unique physical processes. The change in number density is addressed in the following discussion.

\begin{figure}[h]
	\includegraphics[width=\linewidth]{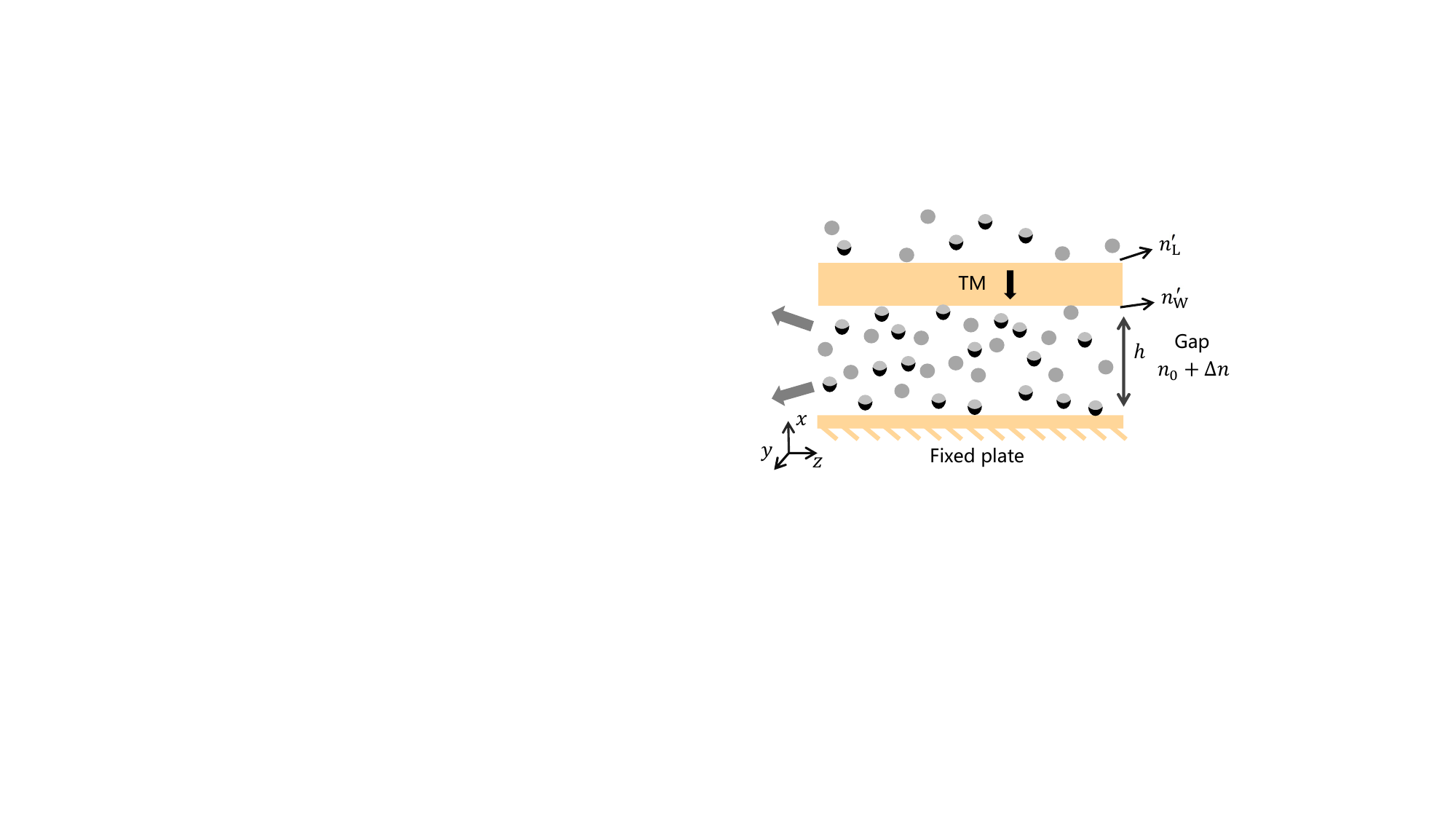}
	\caption{Schematic side view of the bi-plate finite open system. The TM moves downward parallel to the $x$-axis, with a gap spacing denoted as $h$. Gas molecules can exchange with the environment through the gap boundary. The pure light-gray gas molecules are from the fixed plate surface or the environment, while the gas molecules represented by light-gray and black desorb from the TM surface. The color black represents more or less momentum or energy carried compared to the case when the TM is stationary. The effective molecular number density on the windward and leeward sides is denoted as $n'_{\mathrm{W}}$ and $n'_{\mathrm{L}}$, respectively.}
	\label{fig:FTFD}
\end{figure}

In Fig.~\ref{fig:FTFD}, as the TM moves, the gap volume $V$, the total number of molecules $N$ and the number density $n$ undergo continuous changes over time. Considering the total differential of the molecular number density $n(t)=N(t)/V(t)$, and taking the derivative with respect to time, we have
\begin{equation}
	\frac{\mathrm{d}\Delta n}{\mathrm{d}t}=\frac{1}{V_{0}}\frac{\mathrm{d}\Delta N}{\mathrm{d}t}-\frac{N_{0}}{V_{0}^{2}}\frac{\mathrm{d}\Delta V}{\mathrm{d}t},
	\label{eq:fullderivative}
\end{equation}
where $\Delta n$ is the change in molecular number density, $\Delta N(t)$ and $\Delta V(t)$ are the changes in the molecular number and gap volume relative to the equilibrium point. Considering the diffusion process for the molecules to escape into the environment, as shown by the arrows in the left of Fig.~\ref{fig:FTFD}, we deal with the first and second terms on the right side of Eq.~(\ref{eq:fullderivative}) in Appendix~\ref{sec:twoterms}, then under condition $\Delta h\ll h_{0}$, Eq.~(\ref{eq:fullderivative}) is rearranged as
\begin{equation}
	\frac{\mathrm{d}\Delta n}{\mathrm{d}t}=-\frac{\Delta n}{\tau_{0}}-\frac{n_{0}}{h_{0}}\frac{\mathrm{d}h}{\mathrm{d}t},
	\label{eq:fullderivative2}
\end{equation}
where $\tau_{0}$ is diffusion time for one molecule to escape from the gap and $h(t)=h_{0}+B\cos(\omega t)$ is gap spacing in the frequency domain, and the general solution to Eq.~(\ref{eq:fullderivative2}) is
\begin{equation}
	\begin{split}
		\Delta n
		=\frac{n_{0}B\omega\tau_{0}}{h_{0}\sqrt{1+\omega^2\tau_{0}^{2}}}\cos (\omega t+\varphi)+Ce^{-\frac{t}{\tau_{0}}},
	\end{split}
	\label{eq:generalsolution}
\end{equation}
where $\varphi=(\arctan(1/\omega\tau_{0})-\pi)\in(-\pi,-\pi/2)$ is the phase angle, and $C$ is the integration constant. The first term in Eq.~(\ref{eq:generalsolution}) is the steady-state solution, and the second term is the transient solution, which diminishes over time. In the most high-precision gravity-related experiments, the product of detection frequency bands and diffusion times is $\omega\tau_{0}\in(10^{-8}, 10^{-2})$~\cite{PhysRevLett.116.061102,PhysRevLett.116.231101,PhysRevLett.120.061101,Luo_2016,10.1093/nsr/nwx116,PhysRevLett.108.081101, PhysRevLett.116.131101, PhysRevLett.124.051301, PhysRevLett.117.071102, PhysRevLett.122.011102, PhysRevLett.126.211101,naeimi2017global,MOORE20031897,touboul2012champ,10.2514/1.A34326,wu2022global}, and thus $\varphi\approx-\pi/2$. The steady-state part of Eq.~(\ref{eq:generalsolution}) is approximated as:
\begin{equation}
	\begin{split}
		\Delta n
		\approx\frac{n_{0}B\omega\tau_{0}}{h_{0}}\cos (\omega t-\frac{\pi}{2}).\\
	\end{split}
	\label{eq:densityapprox}
\end{equation}
The system's finite openness can be characterized by the diffusion time $\tau_{0}$. A small degree of finite openness means that the number of exchanges between the gap and the environment is small per unit time, equivalent to a large diffusion time $\tau_{0}$, i.e., a long relaxation time for the system, and vice versa. When the gap boundary is closed, the gap spacing change $\Delta h$ is in phase opposition with the density change $\Delta n_{\mathrm{closed}}$, this is consistent with our general understanding. As the gap boundary opens, $\tau_{0}$ decreases, the phase $\varphi$ of $\Delta n$ changes $-\pi\rightarrow -\pi/2$. Interestingly, when $\Delta h$ continues to increase beyond the equilibrium position, $\Delta n_{\mathrm{open}}$ also increases, which may seem counterintuitive. The explanation is that when the $v_{\mathrm{TM}}$ is very small, the difference in number density between the gap and the environment is small. However, when the $v_{\mathrm{TM}}$ is very large, due to the reaction time required for diffusion, the number density in the gap will differ significantly from that in the environment. Furthermore, considering the direction of motion, $\Delta n_{\mathrm{open}}$ should be in phase opposition to $v_{\mathrm{TM}}$.

\section{\label{sec:seperation} Separation of damping coefficients}

The Knudsen hypothesis posits that molecules, upon impacting a surface, undergo adsorption, remain for a brief period, and forget their velocity and direction before impact. During desorption, the speed of emitted molecules follows the Maxwell-Boltzmann distribution, and the emission angles obey cosine law~\cite{PhysRevApplied.19.044005,STECKELMACHER1966561,10.1063/1.1675763,10.1063/5.0011958,luth2001solid,hurlbut1959molecular,STECKELMACHER1966561,10.1063/1.1674619,10.1063/1.1675763,FERES20041541,PhysRevE.77.021202}. According to the cosine law, the probability for a molecule to leave the surface within the solid angle $\mathrm{d}\omega$ is expressed as $\mathrm{d}P=\mathrm{d}\omega\cos\theta/\pi$, where $\theta$ denotes the angle with the surface normal direction. 

The cosine law is initially discovered in experiments using glass surfaces~\cite{knudsen1911molekularstromung,knudsen1917vaporisation,knudsen1917vaporisation}, and subsequent studies on various materials have also verified the same emission rule~\cite{PhysRevD.81.123008,PhysRevApplied.19.044005,STECKELMACHER1966561,10.1063/1.1675763,10.1063/5.0011958,luth2001solid,hurlbut1959molecular,STECKELMACHER1966561,10.1063/1.1674619,10.1063/1.1675763,FERES20041541,PhysRevE.77.021202}. Especially, Ref.~\cite{PhysRevApplied.19.044005} provides a condition for non-elastic collisions in a vacuum ($p_{0}<10^{-4}\ \mathrm{Pa}$) with gold-plated surfaces. 
In realistic applications in the field of MEMS/NEMS, various models considering different boundary conditions are available~\cite{acs.nanolett.1c02237,IMBODEN201489,MinhangBao_2002,BAO20073}, including isothermal and adiabatic, rough and smooth surfaces, diffuse and specular scattering, etc., which are often selected based on experimental data. 

Despite some deviations observed in individual experiments~\cite{10.1063/5.0018726,STECKELMACHER1966561}, it is assumed that the cosine law remains valid for surfaces of general materials. The cosine law and the parallel arrangement of the bi-plate ensure that the gas within the gap adheres to the Maxwell-Boltzmann velocity distribution~\cite{COMSA1985145,lafferty1999foundations,o2023user}. Given the extremely short adsorption time at room temperature (on the order of picoseconds or nanoseconds)~\cite{PhysRevD.84.063007}, much shorter than the motion time of molecules in the gap (on the order of microseconds), it is reasonable to assume a negligible adsorption time.

Next, we will proceed to separate the damping coefficients in the bi-plate and sensor core finite open systems, respectively.

\subsection{\label{subsec:Biplate} Bi-plate finite open system}

As shown in Fig.~\ref{fig:FTFD}, assuming the TM is moving downward, the frequency of collisions with molecules increases on the windward side and decreases on the leeward side, impacting the force experienced by the TM. We will analyze this effect separately for surfaces perpendicular ($x$ surface) and parallel ($y$ or $z$ surface) to the direction of motion.

\emph{Damping force of incident molecules on $x$ surface.}---
Consider a surface element $\mathrm{d}A_{x}$ perpendicular to the $x$-axis. Assuming the TM has motion solely parallel to the $x$-axis, the number of molecular per unit time colliding with the element $\mathrm{d}A_{x}$ on the windward side, with a velocity component $v_{x}$ in the range $v_{x}\sim v_{x}+\mathrm{d}v_{x}$ is given by~\cite{pathria2016statistical,lafferty1999foundations,o2023user}
\begin{equation}
	\mathrm{d}N_{v_{x}}=(n_{0}+\Delta n)(v_{x}-v_{\mathrm{TM}})f(v_{x})\mathrm{d}v_{x}\mathrm{d}A_{x},
	\label{eq:molecularnumber}
\end{equation}
where $f$ is the Maxwell-Boltzmann velocity distribution function, i.e. $f(v_{x})=\left(m/2\pi k_{\mathrm{B}}T\right)^{1/2}\exp\left(-mv_{x}^{2}/2k_{\mathrm{B}}T\right)$. The molecules of $v_{x}>v_{\mathrm{TM}}$ will collide with $\mathrm{d}A_{x}$, so $v_{x}\in (v_{\mathrm{TM}},\infty)$. Taking into account momentum conservation, the force exerted per unit time by collisions on the windward side is
\begin{equation}
	F_{\perp,\mathrm{W}}^{\mathrm{in}}
	=(n_{0}+\Delta n)A_{x}m\int_{v_{\mathrm{TM}}}^{\infty}(v_{x}-v_{\mathrm{TM}})^{2}f(v_{x})\mathrm{d}v_{x},
	\label{eq:Fwindward}
\end{equation}
where $m$ is the molecular mass. Similarly, the force exerted per unit time by collisions on the leeward side is
\begin{equation}
	F_{\perp,\mathrm{L}}^{\mathrm{in}}
	=-n_{0}A_{x}m\int^{v_{\mathrm{TM}}}_{-\infty}(v_{x}-v_{\mathrm{TM}})^{2}f(v_{x})\mathrm{d}v_{x}.
	\label{eq:Fleeward}
\end{equation}

Note that the leeward side is in the external environment, and the number density remains $n_{0}$. From the Eq.~(\ref{eq:densityapprox}), it can be seen that $\Delta n=-n_{0}\tau_{0}v_{\mathrm{TM}}/h_{0}$. Taking into account that in gravity-related experiments, $v_{\mathrm{TM}}$ is very small, only the first-order terms are retained in the subsequent derivation. The total damping force from molecules incident on surfaces perpendicular to the direction of motion is then approximately given by
\begin{equation}
	F_{\perp}^{\mathrm{in}}
	\approx-2A_{x}n_{0}\left(\frac{2mk_{\mathrm{B}}T}{\pi}\right)^{1/2}v_{\mathrm{TM}}-\frac{A_{x}n_{0}\tau_{0}k_{\mathrm{B}}T}{2h_{0}}v_{\mathrm{TM}}.
	\label{eq:Fperpin}
\end{equation}

\emph{Damping force of molecules desorbed from the $x$ surface.}---
Since molecules lose memory after colliding with the surface, the incident and outgoing processes are independent. Therefore, the force exerted by desorbed molecules on the TM can be calculated separately. The cosine law ensures that the incident distribution is the same as the outgoing distribution when taking the TM as the stationary coordinate system~\cite{COMSA1985145,lafferty1999foundations,o2023user}. Due to the identical distributions, the outgoing process is equivalent to the incident process, and Eq.~(\ref{eq:molecularnumber}) can be used directly. As the TM moves, the number of incident molecules will change, so will the number of outgoing molecules. In this scenario, we refer to the outgoing number as the effective incident number.

From Eq.~(\ref{eq:molecularnumber}), after integrating the velocity $v_{x}$, only the variables $n$ and $T$ remain. Under isothermal conditions, $T$ remains constant, so the only variable that can change is $n=n_{0}+\Delta n$. Therefore, the difference between the real incident number and the effective incident number lies in the effective number densities $n'$. By setting the original incident number equal to the outgoing number per unit time, we can write the following identity:
\begin{equation}
	-n_{\mathrm{W}}'\int^{0}_{-\infty}v_{x}f(v_{x})\mathrm{d}v_{x}
	=(n_{0}+\Delta n)\int_{v_{\mathrm{TM}}}^{\infty}(v_{x}-v_{\mathrm{TM}})f(v_{x})\mathrm{d}v_{x},
	\label{eq:inmolecular}
\end{equation}
Considering that as $v_{\mathrm{TM}}\rightarrow 0$, $\mathrm{Erf}\left[\sqrt{m/(2k_{\mathrm{B}}T)}v_{\mathrm{TM}}\right]\rightarrow 0$, and $\exp[-mv_{\mathrm{TM}}^{2}/(2k_{\mathrm{B}}T)]\rightarrow 1$, Eq.~(\ref{eq:inmolecular}) can be solved to obtain the effective number density on the windward side
\begin{equation}
	\begin{split}
		n_{\mathrm{W}}'=(n_{0}+\Delta n)\left[-v_{\mathrm{TM}}\left(\frac{\pi m}{2k_{\mathrm{B}}T}\right)^{1/2}+1\right].
	\end{split}
	\label{eq:nwindward}
\end{equation}
Similarly on the leeward side,
\begin{equation}
	\begin{split}
		n_{\mathrm{L}}'=n_{0}\left[v_{\mathrm{TM}}\left(\frac{\pi m}{2k_{\mathrm{B}}T}\right)^{1/2}+1\right].
	\end{split}
\end{equation}
Thus, the total damping force generated by molecules emitted parallel to the direction of motion is: 
\begin{equation}
	\begin{split}
		F_{\perp}^{\mathrm{out}}
		=-n_{0}A_{x}\left(\frac{\pi mk_{\mathrm{B}}T}{2}\right)^{1/2}v_{\mathrm{TM}}-\frac{A_{x}n_{0}\tau_{0}k_{\mathrm{B}}T}{2h_{0}}v_{\mathrm{TM}}.\\
	\end{split}
	\label{eq:Fperpout}
\end{equation}

\emph{Damping forces on $y$ or $z$ surface.}---
Since the three directions of the Maxwell-Boltzmann velocity distribution are completely independent, we can first use the $v_{y}$ distribution to get the number $\int_{v_{y}}\mathrm{d}N_{y,v_{y}}$ hitting the side, and then use this number to calculate the damping coefficient in the $x$-axis. Since the $z$ direction is perpendicular to the direction of motion of the TM, it does not need to be calculated, and only the velocity distribution of $v_{x}$ in the $x$ direction is considered. Such analysis also applies to surface $z$.

Since $v_{y/z}$ is perpendicular to the direction of motion, the motion of the TM does not affect the velocity distribution and integration limits during calculation. From this, we can get the force per unit area on the side
\begin{equation}
	\mathrm{d}F_{y/z,v_{x}}=m(v_{x}-v_{\mathrm{TM}})f(v_{x})\mathrm{d}v_{x}\mathrm{d}A_{y/z}\int_{v_{y/z}}\mathrm{d}N_{y/z,v_{y/z}}.
	\label{eq:sideforce}
\end{equation}
The outgoing molecule velocities cancel each other out in all directions, so the net force exerted in the $x$ direction is zero. Thus, we obtain the damping force on a single lateral surface:
\begin{equation}
	F_{\parallel ,y/z}=-n_{0}A_{y/z}\left(\frac{mk_{\mathrm{B}}T}{2\pi}\right)^{1/2}v_{\mathrm{TM}}.
	\label{eq:Fparallel}
\end{equation}
Finally, combining Eq.~(\ref{eq:Fperpin}), (\ref{eq:Fperpout}), and (\ref{eq:Fparallel}), we can get the total damping force and then the damping coefficient
\begin{equation}
	\begin{split}
		\beta_{\mathrm{B}}
		=&(4A_{x}+\pi A_{x}+2A_{y}+2A_{z})p_{0}\left(\frac{m}{2\pi k_{\mathrm{B}}T}\right)^{1/2}\\
		&+\frac{A_{x}p_{0}\tau_{0}}{h_{0}},
	\end{split}
	\label{eq:betaB}
\end{equation}
where $p_{0}=n_{0}k_{\mathrm{B}}T$ is the pressure at equilibrium in the gap.

\subsection{\label{subsec:sensitivestructure} Sensor core finite open system}

\begin{figure}[h]
	\includegraphics[width=0.9\linewidth]{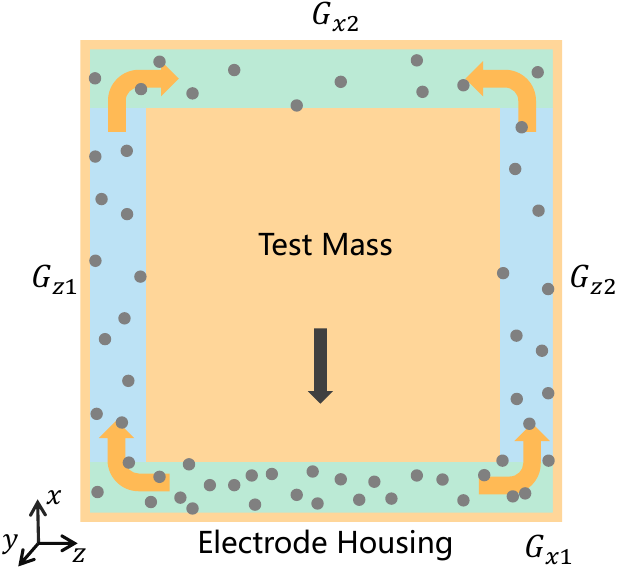}
	\caption{Schematic side view of the sensor core in GRS. The sensor core consists of a TM in the middle and an electrode housing surrounding it, with a gap between them. The gap can be divided into six bi-plate gaps, i.e., bottom gaps $G_{x1}$, $G_{x2}$, and lateral gaps $G_{z1}$, $G_{z2}$, $G_{y1}$, $G_{y2}$, and the last two are not shown.}
	\label{fig:Sensitive_structure}
\end{figure}

As depicted in Fig.~\ref{fig:Sensitive_structure}, the sensor core used in GRSs for space GW detection missions such as LISA and similar projects~\cite{PhysRevLett.116.231101, PhysRevLett.120.061101,Luo_2016,10.1093/nsr/nwx116} is characterized by a cuboid nested structure. This structure is also employed in global gravity field recovery missions albeit with variations in size~\cite{touboul1999accelerometers,touboul2001space,naeimi2017global,MOORE20031897,touboul2012champ,10.2514/1.A34326,wu2022global}, and serves to monitor the real-time position of the TM by connecting to the front-end circuit and apply electrostatic forces to control the TM's behavior. The sensor core forms a closed gap structure between the TM and the electrode housing. However, for a gap on one side, molecules can exchange with other gaps, forming a finite open structure. In Fig.~\ref{fig:Sensitive_structure}, we decompose it into six bi-plate finite open systems, $G_{x1}$, $G_{x2}$, $G_{y1}$, $G_{y2}$, $G_{z1}$, $G_{z2}$, and handle them separately.

\emph{Damping force of incident molecules on $x$ surface.}---
In a manner analogous to the previous subsection \ref{subsec:Biplate}, the force exerted by all molecules hitting the windward face per unit time remains the same as in Eq.~(\ref{eq:Fwindward}). The change in the number density in gap $G_{x1}$ is opposite to the change in number density in gap $G_{x2}$, i.e., $\Delta n_{\mathrm{x2}}=-\Delta n_{\mathrm{x1}}=-\Delta n$. Consequently, the force exerted by all molecules hitting the leeward face per unit time becomes
\begin{equation}
	\begin{split}
		F_{\perp,\mathrm{L}}^{\mathrm{in}}
		=&-(n_{0}-\Delta n)A_{x}m\int^{v_{\mathrm{TM}}}_{-\infty}(v_{x}-v_{\mathrm{TM}})^{2}f(v_{x})\mathrm{d}v_{x}.\\
	\end{split}
	\label{eq:Fleeward2}
\end{equation}
The total damping force of molecules incident perpendicular to the direction of motion is then approximately given by
\begin{equation}
	F_{\bot}^{\mathrm{in}}
	\approx-2A_{x}n_{0}\left(\frac{2mk_{\mathrm{B}}T}{\pi}\right)^{1/2}v_{\mathrm{TM}}-\frac{A_{x}n_{0}\tau_{0}k_{\mathrm{B}}T}{h_{0}}v_{\mathrm{TM}}.
	\label{eq:Finjection}
\end{equation}

\emph{Damping force of molecules desorbed from $x$ surface.}---
The effective molecular number density on the windward face is the same as in Eq.~(\ref{eq:nwindward}), while on the leeward side, the effective number density becomes
\begin{equation}
	n_{\mathrm{L}}'=(n_{0}-\Delta n)\left[v_{\mathrm{TM}}\left(\frac{\pi m}{2k_{\mathrm{B}}T}\right)^{1/2}+1\right],
\end{equation}
The total damping force generated by molecules emitted perpendicular to the direction of motion is given by
\begin{equation}
	\begin{split}
		F_{\perp}^{\mathrm{out}}
		&=-n_{0}A_{x}\left(\frac{\pi mk_{\mathrm{B}}T}{2}\right)^{1/2}v_{\mathrm{TM}}-A_{x}\frac{n_{0}\tau_{0}k_{\mathrm{B}}T}{h_{0}}v_{\mathrm{TM}}.\\
	\end{split}
	\label{eq:Fejection}
\end{equation}

\emph{Damping forces on $y$ or $z$ surface.}---
As the TM moves up and down, the number of molecules in the gaps $G_{x1}$ and $G_{x2}$ changes due to diffusion, causing changes in the number $\Delta N_{\mathrm{side}}$ and number density of molecules in the lateral gap ($G_{y1}$, $G_{y2}$, $G_{z1}$, $G_{z2}$). Molecules diffuse into gap $G_{x1}$ while diffusing out of gap $G_{x2}$, resulting in a density gradient in the $x$-axis direction on the lateral gaps. However, according to Eq.~(\ref{eq:sideforce}), the molecular density on the side is linearly superposed, allowing for non-uniform density distribution on the side. The side force can be reexpressed as
\begin{equation}
	\begin{split}
		F_{\parallel,y/z}
		=&-A_{y/z}n_{0}\left(\frac{mk_{\mathrm{B}}T}{2\pi}\right)^{1/2}v_{\mathrm{TM}}\\
		&-\frac{\Delta N_{\mathrm{side,y/z}}}{h_{y/z}}\left(\frac{mk_{\mathrm{B}}T}{2\pi}\right)^{1/2}v_{\mathrm{TM}}.
	\end{split}
	\label{eq:sideforce2}
\end{equation}
Condsidering the number density changes in the gaps $G_{x1}$ and $G_{x2}$ are $\Delta n_{x1}$ and $\Delta n_{x2}$, and the instantaneous volumes are $V_{x1}=A_{x}(h_{0}+B\cos(\omega t))$ and $V_{x2}=A_{x}(h_{0}-B\cos(\omega t))$, respectively. Thus, we obtain
\begin{equation}
	\begin{split}
		\Delta N_{\mathrm{side}}
		&=-(\Delta n_{x1}V_{x1}+\Delta n_{x2}V_{x2})\\
		&=\frac{4n_{0}B\tau_{0}A_{x}}{h_{0}}\cos(\omega t)v_{\mathrm{TM}},
	\end{split}
	\label{eq:Nside}
\end{equation}
Substituting Eq.~(\ref{eq:Nside}) into Eq.~(\ref{eq:sideforce2}),
\begin{equation}
	F_{\parallel,y/z}
	\approx -A_{y/z}n_{0}\left(\frac{mk_{\mathrm{B}}T}{2\pi}\right)^{1/2}v_{\mathrm{TM}}.
	\label{eq:Fside}
\end{equation}
Unlike the gaps $G_{x1}$ and $G_{x2}$, the fluctuation of number density does not affect the lateral damping force.

Combining Eqs.~(\ref{eq:Finjection}), (\ref{eq:Fejection}), and (\ref{eq:Fside}), we can obtain the total damping force and then the damping coefficient
\begin{equation}
	\begin{split}
		\beta_{\mathrm{S}}
		=&(4A_{x}+\pi A_{x}+2A_{y}+2A_{z})p_{0}\left(\frac{m}{2\pi k_{\mathrm{B}}T}\right)^{1/2}\\
		&+\frac{2A_{x}p_{0}\tau_{0}}{h_{0}}.
	\end{split}
	\label{eq:betaS}
\end{equation}
The first term in Eqs.~(\ref{eq:betaB}) and (\ref{eq:betaS}) is consistent with the free damping in the infinite gas volume~\cite{CAVALLERI20103365}, while the second term is consistent with the proximity damping in the isothermal piston model~\cite{SUIJLEN2009171,PhysRevD.84.063007}. The first term can be regarded as a basic damping that always exists regardless of the environment, whereas the second term represents damping induced by molecular diffusion. We refer to them as base damping and diffusion damping, respectively. 

The models described above are denoted as finite open models. Finally, we achieve the natural separation of damping coefficients within the finite open systems.

\section{\label{sec:energy} Separation and sources of damping from an energy perspective}

As shown in Fig.~\ref{fig:FTFD}, when the molecules desorb from the moving TM surface, in addition to the velocity given by the Maxwell-Boltzmann distribution, there is also a background translational velocity of the TM. In other words, the TM only affects the energy of the molecules in translational degrees of freedom.

Using subscripts $\mathrm{u}$ for gas molecules desorbed from the TM surface, and $\mathrm{d}$ for those desorbed from the fixed plate below, with the total gas molecules indexed as $i$, the total energy can be expressed as
\begin{equation}
	\begin{split}
		E
		&=\frac{1}{2}\sum_{i}mv_{i}^{2}=\frac{1}{2}\sum_{\mathrm{d}}mv_{\mathrm{d}}^{2}+\frac{1}{2}\sum_{\mathrm{u}}mv_{\mathrm{u}}^{2}\\
		&\approx\frac{1}{2}\sum_{i}mv_{i}'^{2}+\frac{1}{2}\sum_{\mathrm{u}}mv_{\mathrm{TM}}^{2}+\sum_{\mathrm{u}}m\bm{v}_{\mathrm{u}}'\cdot\bm{v}_{\mathrm{TM}},\\
	\end{split}\label{eq:internalenergy}
\end{equation}
where $\bm{v}_{i}'$ and $\bm{v}_{\mathrm{u}}'=\bm{v}_{\mathrm{u}}-\bm{v}_{\mathrm{TM}}$ represent the velocities of molecules in the center-of-mass frame assuming that the TM does not move. The first term in Eq.~(\ref{eq:internalenergy}), $\sum_{i}mv_{i}'^{2}/2\approx\sum_{\mathrm{d}}mv_{\mathrm{d}}^{2}/2+\sum_{\mathrm{u}}mv_{\mathrm{u}}'^{2}/2$, assumes that the number $N_{\mathrm{u}}$ of molecules desorbed from the TM surface and the number $N_{\mathrm{d}}$ desorbed from the bottom fixed plate are approximately equal, $N_{\mathrm{u}}\approx N_{\mathrm{d}}$. Molecules desorb from the TM surface and emit to the fixed plate, and then back to the TM surface, this process takes a very short time, $\Delta t\sim 10^{-5}\ \mathrm{s}$, while the TM's velocity is very slow, $v_{\mathrm{TM}}<10^{-8} \mathrm{m/s}$. In other words, during this time, $N_{\mathrm{d}}(t_{0}+\Delta t)\approx N_{\mathrm{u}}(t_{0})$, so this approximation is reasonable. The total number $N(t)=N_{\mathrm{u}}+N_{\mathrm{d}}$ changes due to the diffusion of gas molecules.

Let $v_{\mathrm{u}x}'$ be the projection of $\bm{v}_{\mathrm{u}}'$ in the direction of the TM motion, its probability density distribution is $2f(v_{\mathrm{u}x}')$. Therefore, the third term can be re-expressed as
\begin{equation}
	\sum_{\mathrm{u}}m_{\mathrm{u}}\bm{v}_{\mathrm{u}}'\cdot\bm{v}_{\mathrm{TM}}=N_{\mathrm{u}}\left(\frac{2mk_{\mathrm{B}}T}{\pi}\right)^{1/2}v_{\mathrm{TM}}.
\end{equation}
Considering the Knudsen hypothesis, the Eq.~(\ref{eq:internalenergy}) can be re-expressed as
\begin{equation}
	\begin{split}
		E(t)=\frac{3}{2}N(t)k_{\mathrm{B}}T+\frac{1}{4}N(t)mv_{\mathrm{TM}}^{2}+N(t)\left(\frac{mk_{\mathrm{B}}T}{2\pi}\right)^{1/2}v_{\mathrm{TM}}.
	\end{split}
	\label{eq:totalenergy}
\end{equation}
The first term is consistent with the isothermal internal energy of the gas in an equilibrium state, and the second and third terms are related to the translational velocity of the TM, representing additional energy introduced by non-equilibrium processes. This result is different from the König's theorem~\cite{koenig1751universali}, which states that the energy of the system can be separated into the translational kinetic energy of the center-of-mass and the relative kinetic energy within the center-of-mass system. In the Maxwell-Boltzmann distribution, the overall translational velocity of the gas is zero. In other words, when we refer to temperature, this means that the system is discussed in the center-of-mass system. Thus, it is inaccurate to use isothermal gas or effective temperature assumptions in this system. In the free molecular flow regime, molecules do not collide, thus energy cannot be evenly distributed among the degrees of freedom~\cite{maxwell2003illustrations}, except for the translational degrees of freedom related to the TM. 

According to analytical mechanics, the damping coefficient means the dissipation of energy, and there are three pathways of energy dissipation in the bi-plate gap. When molecules are incident on the TM surfaces, the incident damping forces, i.e. Eqs.~(\ref{eq:Fperpin}) and (\ref{eq:Finjection}) correspond to the energy dissipation of the TM. In Fig.~\ref{fig:FTFD}, due to the motion of the TM, the molecules desorbed from the TM will dissipate more kinetic energy of the TM to the fixed plate. This part of the energy results in the addition of the last two terms in Eq.~(\ref{eq:totalenergy}). For the bi-plate size at the bottom gaps of the sensor core for LISA mission, only about 9\% molecules desorbed from the TM (see Appendix~\ref{sec:outrate}) leave the gap directly. Similarly, the ratio from the environment directly to TM is the same. The ratio is proportional to the energy dissipated, so the majority of the dissipated energy flows to the fixed plate and the TM, with a very small amount flowing to the environment.

Compared to base damping, diffusion damping accounts for changes in number density resulting from the diffusion process. Since the change in number density is in counter-phase with TM's velocity, it is equivalent to increasing the number of incident and outgoing molecules per unit time, which in turn amplifies damping force or energy dissipation. Consequently, the dissipation flow from the gap to the environment as noticed in Ref.~\cite{PhysRevD.84.063007} is not a leading term to the diffusion damping of the energy dissipation. Instead, as mentioned above, this energy is primarily dissipated into the TM and the fixed plate. In isothermal piston models, the utilization of the ideal gas state equation implies that compression is a quasi-static process, wherein only the diffusion damping resulting from changes in number density is considered. As a result, its internal energy corresponds solely to the first term in Eq.~\ref{eq:totalenergy}, excluding the last two terms associated with the damping of the non-equilibrium part.

\section{\label{sec:application} Applications}

The investigation of residual gas noise holds significant importance in gravity-related experiments. Residual gas damping noise typically determines the limit of sensitivities in many high-precision gravity-related experiments. In the subsequent sections, we apply the sensor core and bi-plate finite open models to space gravitational wave detection and the test of the gravitational inverse-square law, respectively.

\subsection{\label{subsec:sensitive} Space gravitational wave detection}

The discovery of GWs has positioned it as a new messenger for observing the universe. LISA is a spaceborne GW detection mission proposed by the European Space Agency~\cite{PhysRevLett.116.231101,PhysRevLett.120.061101,amaroseoane2017laser}. To investigate the residual gas damping noise used in such missions, the University of Trento (UTN) employs a torsion balance to measure the decay curves in the GRSs~\cite{PhysRevLett.103.140601}.

\begin{figure}[h]
	\centering
	\includegraphics[width=0.9\linewidth]{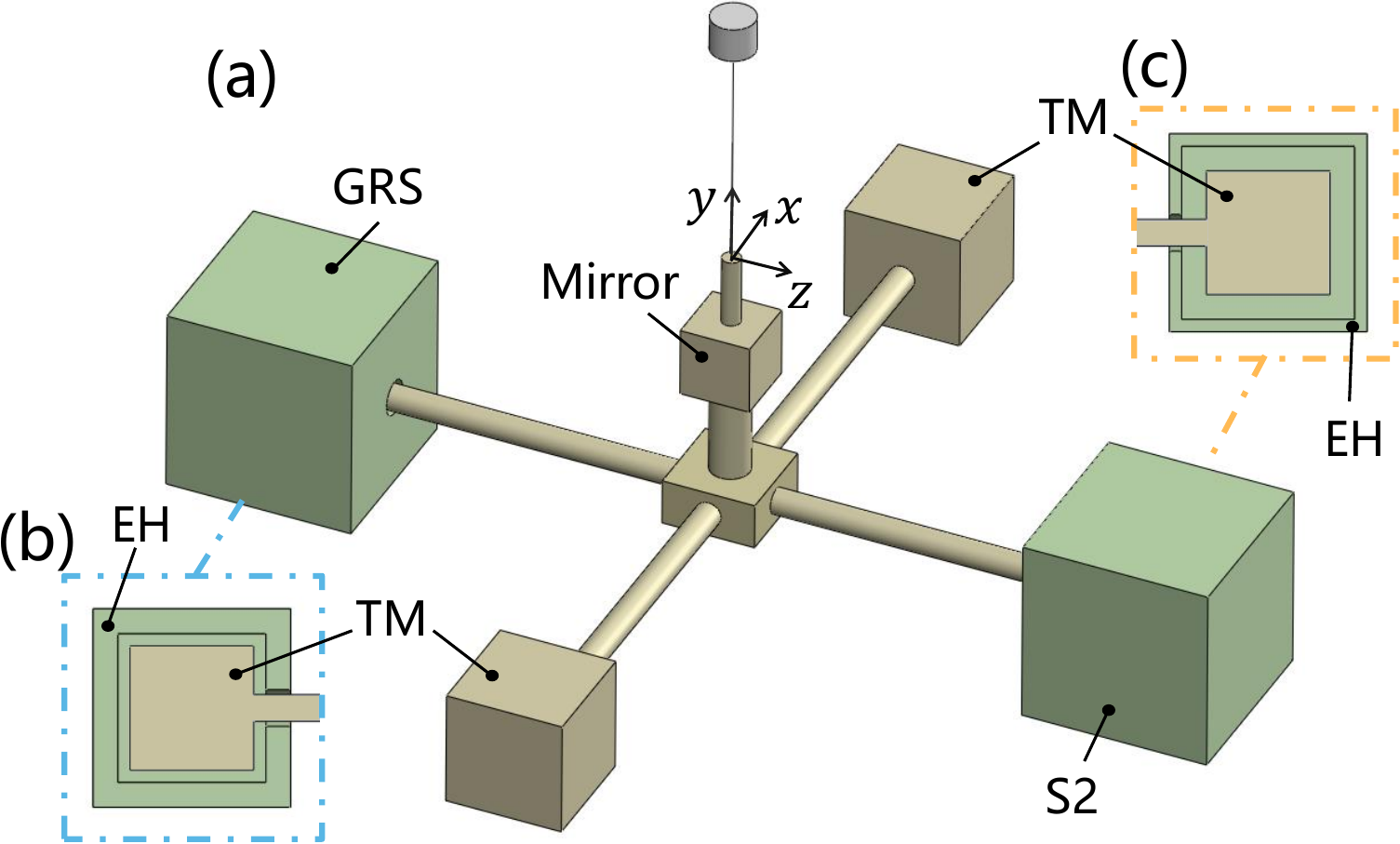}
	\caption{Schematic drawing of the 4TMs torsion balance experiment setup (not to scale). (a) A cross-shaped structure is suspended by a wire, with four hollow cubic TMs attached at each end. The side length of each TM is $s=46\ \mathrm{mm}$, and the pendulum arm length is $r=0.1065\ \mathrm{m}$. (b) The gap spacings between the GRS's TM and the electrode housing (EH) in the $x$, $y$, and $z$ axes are $4.0\ \mathrm{mm}$, $2.9\ \mathrm{mm}$, and $3.5\ \mathrm{mm}$, respectively. (c) The gap spacings between the S2's TM and the EH in the $x$, $y$, and $z$ axes are $8.0\ \mathrm{mm}$, $6.0\ \mathrm{mm}$, and $8.0\ \mathrm{mm}$, respectively~\cite{PhysRevLett.103.140601}.}
	\label{fig:4TMssetup}
\end{figure} 

\begin{figure}[h]
	\includegraphics[width=0.95\linewidth]{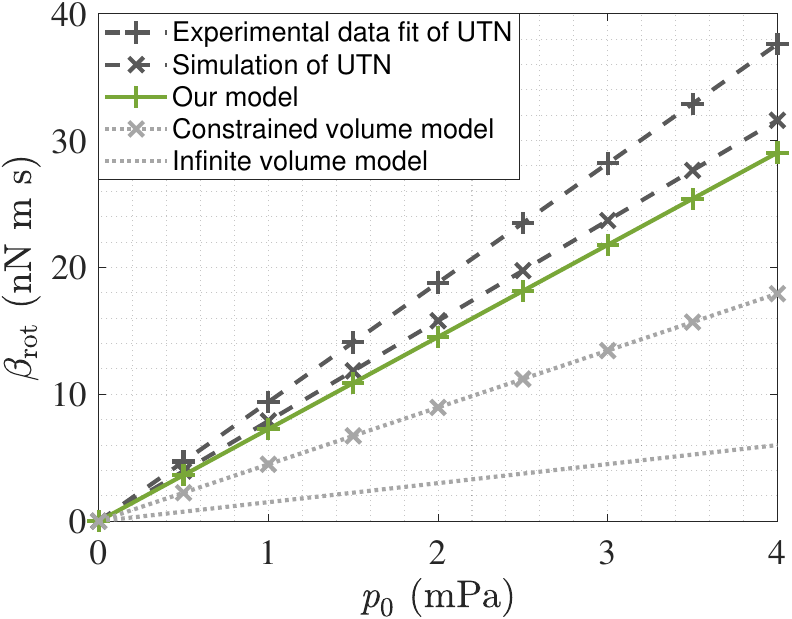}
	\caption{Comparison of different rotational damping coefficients in the 4TM torsion balance experiment. The black dashed line ($+$) and black dashed line ($\times$) represent the experimental measured values and simulation values from the UTN, respectively~\cite{PhysRevLett.103.140601}. The green solid line ($+$) represents the theoretical values from Eq.~(\ref{eq:torsiondamping}), i.e., our finite open model in this paper, and the short dashed gray line ($\times$) represents the theoretical values from the constrained volume model~\cite{Mao_2023}. The short dashed gray line represents the theoretical result of damping coefficients in an infinite volume~\cite{CAVALLERI20103365}.}
	\label{fig:torsiondamping}
\end{figure}

The 4TMs torsion balance experiment setup is shown in Fig.~\ref{fig:4TMssetup}. The experimental temperature is $293\ \mathrm{K}$, and the molecular mass utilized in the simulation is $30\ \mathrm{u}$. By measuring the amplitude decay time during free pendulum motion, the torsional damping coefficients can be obtained. The experimental results indicate that the residual gas damping coefficient is much larger than the theoretical prediction of infinite volume damping. However, it is in good agreement with UTN's simulation predictions, as shown in Fig.~\ref{fig:torsiondamping}.

Since the torsion balance measures the torsional damping coefficient, the translational damping coefficient $\beta_{\mathrm{S}}$ of the finite open model is needed to convert to the 4TM rotational damping coefficient $\beta_{\mathrm{rot}}^{\mathrm{4TM}}=r^{2}\beta_{\mathrm{tran}}+\beta_{\mathrm{rot}}$. The total rotational damping coefficient is given by
\begin{equation}
	\begin{split}
		\beta_{\mathrm{rot}}^{\mathrm{4TM}}
		=&r^{2}p_{0}s^{2}\left[\frac{2\tau_{\mathrm{side}}(h_{x})}{h_{x}}+\left(\frac{32m}{\pi k_{\mathrm{B}}T}\right)^{1/2}\left(1+\frac{\pi}{8}\right)\right]\\
		&+\frac{1}{6}p_{0}s^{4}\left[\frac{\tau_{\mathrm{side}}(h_{x})}{h_{x}}+\frac{\tau_{\mathrm{side}}(h_{z})}{h_{z}}\right]\\
		&+p_{0}s^{4}\left(\frac{2m}{\pi k_{\mathrm{B}}T}\right)^{1/2}\left(1+\frac{\pi}{12}\right).
	\end{split}
	\label{eq:torsiondamping}
\end{equation}

In Fig.~\ref{fig:torsiondamping}, the result of the finite open model demonstrates a 62\% improvement compared to the constrained volume model~\cite{Mao_2023}, and exhibits an 8\% difference from the UTN's simulation data. The latter utilizes the diffusion time under bi-plate structures, but the electrode housing alters the diffusion conditions at the bi-plate boundaries, subsequently changing the diffusion time under the bi-plates. This alteration increases the diffusion time. Finally, we substitute the simulation result into the finite open model, with $\tau_{\mathrm{side}}(h_{x})=3.22\times 10^{-4}\ \mathrm{s}$ and $\tau_{\mathrm{side}}(h_{z})=3.02\times 10^{-4}\ \mathrm{s}$, while $\tau_{\mathrm{cuboid}}=1.31\times 10^{-4}\ \mathrm{s}$. The 8\% discrepancy may come from the gradient of molecular number density in the lateral gap ($G_{y1}$, $G_{y2}$, $G_{z1}$, $G_{z2}$) in the $x$-axis. This might lead to the deviation, i.e., the $N_{0}/\tau_{0}$ term in Eq.~(\ref{eq:diffusionequation}) is not accurate.

The results in Fig.~\ref{fig:torsiondamping} demonstrate the effectiveness of the finite open model. Ultimately, we apply the finite open model to the analysis of the residual gas damping noise in the GRS's sensitive axis for the LISA space GW detection mission~\cite{PhysRevLett.103.140601} and other LISA-like space GW detection missions, such as Taiji~\cite{10.1093/nsr/nwx116}, TianQin~\cite{Luo_2016}. This noise largely determines the detection sensitivity of GRSs in the primary frequency range. For the $2\ \mathrm{kg}$ TM moving along the sensitive axis ($x$-axis), we can use Eqs.~(\ref{eq:fluctuationdamping}) and (\ref{eq:torsiondamping}) to estimate the residual gas acceleration noise, that is
\begin{equation}
	\begin{split}
		S^{1/2}_{a}
		=&1.23\times 10^{-15}\ \mathrm{m\ s^{-2}Hz^{-1/2}}\\
		&\times\left(\frac{p}{10^{-6}\ \mathrm{Pa}}\right)^{1/2}\left(\frac{m}{30\ \mathrm{u}}\right)^{1/4}\left(\frac{T}{293\ \mathrm{K}}\right)^{1/4}.
	\end{split}
\end{equation}
Under the same conditions, compared to Ref.~\cite{Mao_2023}, our result is closer to the simulation result $S^{1/2}_{a} = 1.3 \times 10^{-15}\ \mathrm{m\ s^{-2}Hz^{-1/2}}$ from Ref.~\cite{PhysRevLett.103.140601}, with only a 5\% difference.

The validity of the finite open model also corrects the previous view on the diffusion time in the sensor core. Ref.~\cite{PhysRevLett.103.140601} states that the diffusion time is the average time for molecules to diffuse from gap $G_{x1}$ to $G_{x2}$. However, as shown in Eq.~(\ref{eq:Nside}), we found that the molecular number changes from $G_{x1}$ and $G_{x2}$ have a negligible second-order effect on the $y$/$z$ surfaces, and the diffusion time can be approximated as the time to diffuse just from the bottom gap ($G_{x1}$ or $G_{x2}$) to the lateral gap ($G_{y1}$, $G_{y2}$, $G_{z1}$, $G_{z2}$).

\subsection{\label{subsec:noise} Test of the gravitational inverse-square law}

\begin{figure}[h]
	\centering
	\includegraphics[width=0.9\linewidth]{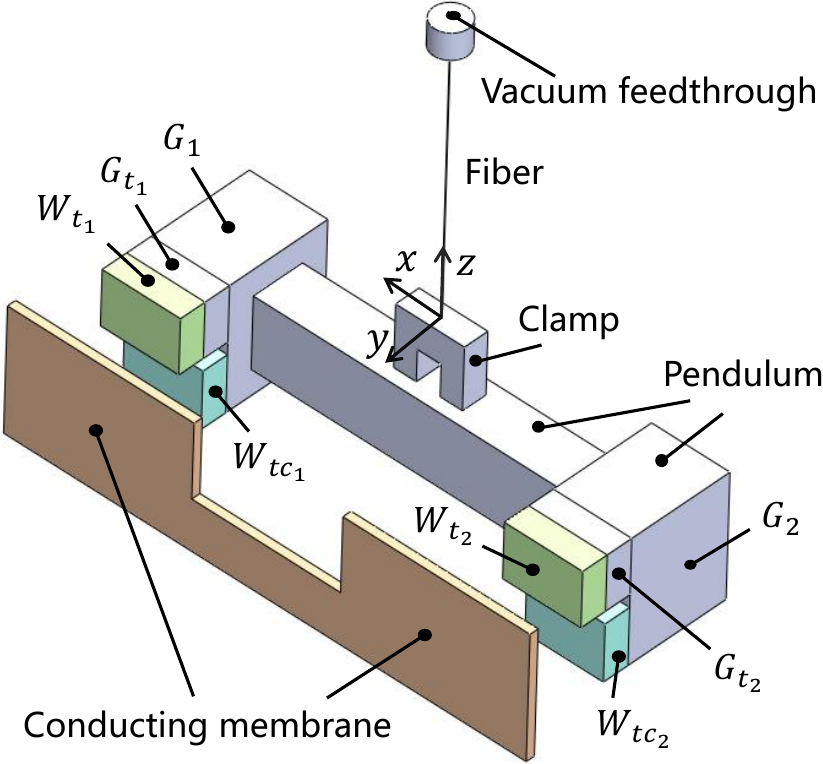}
	\caption{Schematic drawing of the test of the ISL experiment setup (not to scale). An I-shaped pendulum is suspended from the bottom of the torsion balance through tungsten wire, and the pendulum faces the position of the attractor. The middle part of the pendulum is a glass block (M), and two glass bases ($G_{1}$, $G_{2}$) are symmetrically installed at both ends. Two glass substrates ($Gt_{1}$, $Gt_{2}$), two tungsten TMs ($Wt_{1}$, $Wt_{2}$), and two gravitational compensation pieces ($Wtc_{1}$, $Wtc_{2}$) are glued to the glass bases opposite the attractor. A special glass clamp is installed on the top of the middle glass block, it can fix the suspended tungsten wire to the center of the pendulum. On the left side of the conducting membrane, there is an attractor not shown here. The attractor consists of 8 tungsten source masses and 8 tungsten compensation masses, arranged alternately on a rotatable glass disk~\cite{PhysRevLett.116.131101}.}
	\label{fig:ISLsetup}
\end{figure}

To reconcile general relativity and the standard model, string theory or M-theory predicts deviations from the ISL at short distances~\cite{doi:10.1146/annurev.nucl.53.041002.110503,ADELBERGER2009102}. Huazhong University of Science and Technology (HUST) designs a torsion balance and a eightfold azimuthal symmetric attractor experimental platform, and tests the ISL at short distances by dually modulating the signal of interest and the gravity calibration signal~\cite{PhysRevLett.108.081101,PhysRevLett.116.131101,PhysRevLett.124.051301,PhysRevLett.117.071102,PhysRevLett.122.011102,PhysRevLett.126.211101}. 

The experimental setup of the test of the ISL is illustrated in Fig.~\ref{fig:ISLsetup}. To facilitate the experimental design of the gap spacing between the pendulum and the conducting membrane, it is necessary to assess the influence of residual gas damping noise with varying gap spacings. To obtain the constraints on gap spacing $h_{\mathrm{c}}$ under different gas pressures $p_{0}$~\cite{Ke_2022}, the gas damping noise is equated to the torsion balance's internal damping thermal noise. The internal thermal noise of the torsion balance is $S_{\mathrm{th}}(\omega)=4k_{\mathrm{B}}Tk/(\omega Q)\approx 3.61\times 10^{-30}\ \mathrm{N^{2}\ m^{2}\ Hz^{-1}}$. Here, $Q$ is the quality factor of the torsion balance, $\omega$ is the resonant frequency, and $k$ is the spring constant. According to the bi-plate finite open model, the fluctuation torque noise power spectral density $S_{\mathrm{rot,M}}$ of the middle glass block, and $S_{\mathrm{rot},G,Wt}$ of the glass blocks ($G_{1}$, $G_{2}$), TMs ($Wt_{1}$, $Wt_{2}$) and gravitational compensation pieces ($Wtc_{1}$, $Wtc_{2}$) at both ends can be obtained in Appendix~\ref{sec:ISLnoisespectrum}. From this, the relation
\begin{equation}
	\begin{split}
		p_{0}=f(h_{\mathrm{c}})
		=\frac{S_{\mathrm{th}}}{(S_{\mathrm{rot,M}}+S_{\mathrm{rot},G,Wt})/p_{0}},
	\end{split}
\end{equation}
is obtained, and consequently
\begin{equation}
	\begin{split}
		h_{\mathrm{c}}=f^{-1}(p_{0}).
	\end{split}
	\label{eq:ISLapplication}
\end{equation}

\begin{figure}[h]
	\centering
	\includegraphics[width=0.95\linewidth]{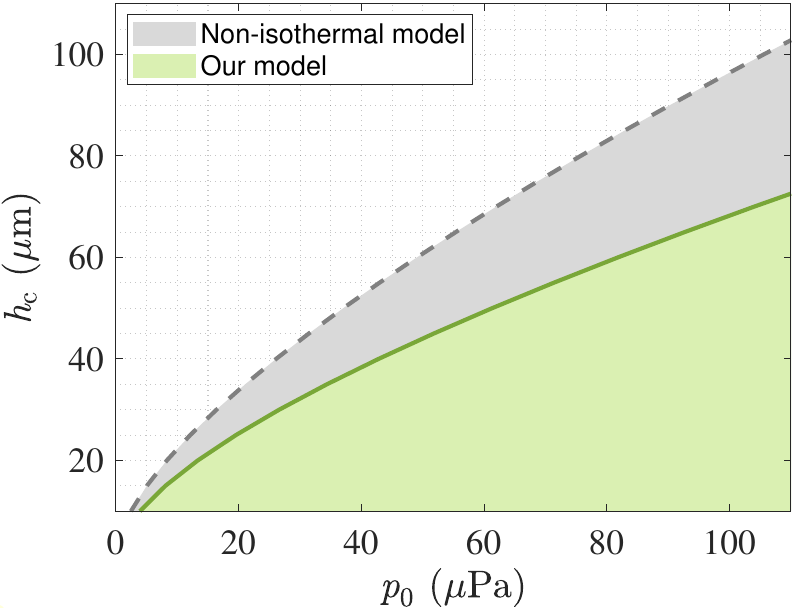}
	\caption{Comparison of gap spacing constraints under different pressures between the finite open model and the non-isothermal model. The shaded area below the green solid line represents the constraints given by Eq.~(\ref{eq:ISLapplication}), i.e., our bi-plate finite open model in this paper, and the shaded area below the gray dashed line represents the constraints given by the non-isothermal model~\cite{Ke_2022}.}
	\label{fig:ISL}
\end{figure}

Given temperature $T=297\ \mathrm{K}$, molecular mass $19.1\ \mathrm{u}$, and thermal adjustment coefficient $\sigma=1$, the results are shown in Fig.~\ref{fig:ISL}. Due to the very small gap spacing, $\sim 100\ \mathrm{\mu m}$, the diffusion damping is magnified, and the base damping is almost negligible. In the range of 10 to 100 $\mathrm{\mu Pa}$, the constraints on the gap spacing $h_{\mathrm{c}}$ obtained by the finite open model are lower by about $28\%$ compared to the non-isothermal model. This suggests that the additional temperature effect in the non-isothermal model is almost negligible. These results are crucial for reducing the test length scale
$\lambda$ in ISL experiments, providing valuable insights into the constraints imposed by residual gas damping noise on experimental parameters.

\section{\label{sec:conclusion} Discussions and Conclusions}

We separate the damping coefficients for two finite open systems, i.e., the bi-plate system and the sensor core system, into base damping and diffusion damping through rigorous derivation. This effectively elucidates the relationship between the free damping and the proximity damping. This separation needs to meet the condition $\omega\tau_{0}\rightarrow 0$, which holds for most  high-precision gravity-related experiments. In our derivation, the key lies in how to deal with the change in molecular number density. The change is so small compared to the overall density that it can be almost ignored~\cite{SUIJLEN2009171}, as evidenced in the LISA's GRSs ($\Delta n/n_{0} < 10^{-11}$). However, detailed derivation shows that the results corresponding to such a small amount can not be ignored, which renders high-precision measurements truly distinctive. We address the impact of changes in number density within the bottom gaps on the lateral gaps, which also reveals the possibilities for analyzing cross-talk effects in residual gas noise between the sensitive axis and other axes.

The diffusion time $\tau$ is used to quantitatively characterize the degree of finite openness in the system, which is different from the constrained volume. It is general and can account for the effects of different shapes of the structure around the TM. However, in a sensor core with lateral walls, an increase in constrained volume does not necessarily equate to a reduction in diffusion time. Interestingly, as $\tau\rightarrow 0$, i.e., the system is completely open, the diffusion damping disappears, and the damping becomes consistent with the free damping in an infinite volume. Conversely, as $\tau\rightarrow \infty$, i.e., the system is completely closed, in contrast to the zero damping of the isothermal piston model~\cite{SUIJLEN2009171,PhysRevD.84.063007}, there still exists base damping independent of gap spacing. This result is reflected in the lower limit of the power spectral density shown in Ref.~\cite{PhysRevD.84.063007}. In brief, the concept of finite openness can characterize damping more comprehensively and accurately.

In terms of the physical picture, the previous analyses of damping in an infinite volume and a constrained volume are disjointed. The finite open models presented in this paper provide a complete microscopic picture that includes both types of damping. This allows us to point out three pathways of energy dissipation of the TM from the microscopic level in the bi-plate gap, that is: energy absorbed by the TM, assimilated by the fixed plate, and a very small portion directly diffused into the environment. 

In practical applications, for space GW detection, our model predicts a damping coefficient for the 4TM torsion balance experiment 62\% higher than previous analyses, and is the closest to the experimental and simulation data to the best of our knowledge. For the LISA mission, our theoretical estimation of the residual gas acceleration noise along the sensitive axis aligns most closely with simulation results, differing only by 5\%. These results validate the effectiveness of the finite open models. And these results are also applicable to other LISA-like space GW detection missions, such as Taiji~\cite{10.1093/nsr/nwx116}, TianQin~\cite{Luo_2016}. In the test of the ISL, our model reduces the constraints on the gap spacing between the TM and the conducting membrane by about $28\%$, compared with the non-isothermal model. And this result may further help to reduce the test length scale in ISL experiments.

Furthermore, our theory can also be applied to other high-precision measurements, such as ultra-sensitive space accelerometers~\cite{touboul1999accelerometers, touboul2001space, Luo_2020,taiji2021china}, measurements of the gravitational constant $G$~\cite{li2018measurements,RevModPhys.93.025010,rosi2014precision}, MEMS/NEMS accelerometers~\cite{fan2019graphene,acs.nanolett.9b01759} and Casimir force measurements~\cite{PhysRevLett.88.041804,fong2019phonon}.

\begin{acknowledgments}
	We thank Profs. Yun-Kau Lau and Lamberto Rondoni for their inspiring discussions and advice, as well as Drs. Zuolei Wang and Da Fan for their help. This work is supported by the National Key Research and Development Program of China under Grant No. 2020YFC2200601, the National Natural Science Foundation of China under Grants No. 12175090, No. 12305045 and No. 12247101, and the 111 Project under Grant No. B20063.
\end{acknowledgments}


\appendix

\section{\label{sec:twoterms} The Two Terms in the Equation (\ref{eq:fullderivative})}

\emph{The first term.}---
Suppose the average diffusion time for one molecule to escape from the gap is $\tau(h)$, and $N/\tau$ is the number of escaping molecules per unit time, then
\begin{equation}
	\mathrm{d}\Delta N=-\frac{N(t)}{\tau(h)}\mathrm{d}t+\frac{N_{0}}{\tau_{0}}\mathrm{d}t,
	\label{eq:diffusionequation}
\end{equation}
where $\Delta N$ is the net number of diffusing molecules, $N_{0}$ and $\tau_{0}$ are at the equilibrium point. Equation~(\ref{eq:diffusionequation}) is also known as the Fick rule~\cite{Ke_2022}. Considering gap spacing change $\Delta h\ll h_{0}$, we can have the approximation $\tau(h)\approx \tau_{0}$. In this case, dividing Eq.~(\ref{eq:diffusionequation}) by $\mathrm{d}t$ and $V_{0}$ gives
\begin{equation}
	\frac{1}{V_{0}}\frac{\mathrm{d}\Delta N}{\mathrm{d}t}=-\frac{n(t)}{\tau_{0}}+\frac{n_{0}}{\tau_{0}},
	\label{eq:firstitem2}
\end{equation}
where $n_{0}=N_{0}/V_{0}$. In addition, considering $\Delta V(t)\ll V_{0}$, the approximation $N(t)/V_{0}\approx n(t)$ is made. 

\emph{The second term.}---
Assuming the TM vibrates near the equilibrium point, in the frequency domain, the displacement relative to the equilibrium point is $\Delta h(t)=B\cos(\omega t)$, and the instantaneous volume of the gap is $V(t)=A_{x}(h_{0}+B\cos(\omega t))=A_{x}h(t)$, where $A_{x}$ is the surface area of the TM perpendicular to the $x$-axis. Thus, the second term on the right side of Eq.~(\ref{eq:fullderivative}) can be written as
\begin{equation}
	\begin{split}
		-\frac{N_{0}}{V_{0}^{2}}\frac{\mathrm{d}\Delta V}{\mathrm{d}t}
		=-n_{0}\frac{\mathrm{d}}{\mathrm{d}t}\left(\frac{A_{x}h}{V_{0}}\right)
		=-\frac{n_{0}}{h_{0}}\frac{\mathrm{d}h}{\mathrm{d}t}.
	\end{split}
	\label{eq:seconditem}
\end{equation}

\section{\label{sec:outrate} The proportion of Molecules Escaping Directly from the Test Mass Surface to the Environment}

Molecules undergo an average of $\mathcal{N}$ collisions inside the gap before leaving. The last time the molecules escape directly from the TM surface, the proportion is $1/2$, so the proportion of molecules escaping directly from the TM surface to the environment is approximately given by
\begin{equation}
	P_{\mathrm{out}}\approx\frac{1}{2}\frac{1}{\frac{\mathcal{N}}{2}}\approx0.09,
\end{equation}
where we use the conclusion from Ref.~\cite{PhysRevD.84.063007}. For the size of sensor core for LISA mission, $\mathcal{N}$ is approximately given by
\begin{equation}
	\mathrm{\mathcal{N}}\approx\frac{R^{2}}{h^{2}\ln\left[1+\left(\frac{R}{h}\right)^{2}\right]}\approx 11,
\end{equation}
where the radius $R$ is obtained by assuming that the area of the square is consistent with the area of the equivalent circle. The proportion of molecules escaping directly from the TM surface to the environment is 0.09, which is a small fraction.

\section{\label{sec:ISLnoisespectrum} Fluctuating Force Power Spectral Density in ISL Experiment}

According to the theoretical model of the bi-plate finite open system, the fluctuating torque noise $S_{\mathrm{rot,M}}$ for the middle glass block can be obtained as follows:
\begin{equation}
	\begin{split}
		S_{\mathrm{rot,M}}
		=&\frac{s_{\mathrm{M},x}^{3}s_{\mathrm{M},z}}{6}\left(\frac{32mk_{\mathrm{B}}T}{\pi}\right)^{1/2}\left(1+\frac{\pi}{4}\right)p_{0}\\
		&+\left(\frac{s_{\mathrm{M},x}^{3}s_{\mathrm{M},z}}{2}+\frac{s_{\mathrm{M},x}^{3}s_{\mathrm{M},y}}{6}+\frac{s_{\mathrm{M},y}^{3}s_{\mathrm{M},x}}{6}\right)\\
		&\times\left(\frac{8mk_{\mathrm{B}}T}{\pi}\right)^{1/2}p_{0}.
	\end{split}
\end{equation}

The torque noise $S_{\mathrm{rot},G,Wt}$ for the glass blocks ($G_{1}$, $G_{2}$), the TMs ($Wt_{1}$, $Wt_{2}$), and gravitational compensation pieces ($Wtc_{1}$, $Wtc_{2}$) at both ends is given by:
\begin{equation}
	S_{\mathrm{rot},G,Wt}=8k_{\mathrm{B}}T\mathcal{D}l^{2}p_{0},
\end{equation}
where
\begin{equation}
	\begin{split}
		\mathcal{D}=&(4s_{G,x}s_{G,z}+\pi s_{G,x}s_{G,z}+2s_{G,y}s_{G,z}+2s_{G,x}s_{G,y}\\
		&-s_{\mathrm{M},z}s_{\mathrm{M},y})\left(\frac{m}{2\pi k_{\mathrm{B}}T}\right)^{1/2}+\frac{s_{Wt,x}s_{Wt,z}\tau_{0,Wt}}{h_{\mathrm{c}}}\\
		&+\frac{s_{Wtc,x}s_{Wtc,z}\tau_{0,Wtc}}{h_{\mathrm{c}}+(s_{Gt,y}+s_{Wt,y}-s_{Wtc,y})},
	\end{split}
\end{equation}
where, the arm length from the suspension point to the center of the TM is $l=(s_{G,x}+s_{\mathrm{M},x})/2=38.0605\ \mathrm{mm}$, and the dimensions are given as $s_{\mathrm{M},x}\times s_{\mathrm{M},y}\times s_{\mathrm{M},z}=61.491\times 8.000\times 12.000\ \mathrm{mm^{3}}$, $s_{G,x}\times s_{G,y}\times s_{G,z}=14.630\times 19.756\times 27.138\ \mathrm{mm^{3}}$, $s_{Gt,x}\times s_{Gt,y}\times s_{Gt,z}=14.630\times 0.486\times 12.003\ \mathrm{mm^{3}}$, $s_{Wt,x}\times s_{Wt,y}\times s_{Wt,z}=14.630\times 0.200\times 12.003\ \mathrm{mm^{3}}$, $s_{Wtc,x}\times s_{Wtc,y}\times s_{Wtc,z}=14.630\times 0.606\times 15.139\ \mathrm{mm^{3}}$. The diffusion time $\tau_{0,Wt}$ and $\tau_{0,Wtc}$ use $\tau_{\mathrm{cuboid}}$ in Appendix~\ref{sec:diffusiontime}. The parameters here may differ somewhat from those in Refs.~\cite{PhysRevLett.116.131101,Ke_2020,Ke_2022}. The parameters in this paper are for reference only.

Considering that the base damping here is almost negligible, theoretically, when the diffusion time used in the non-isothermal model is the same as the diffusion time formula used in the bi-plate finite open model, the ratio of the power spectral density of fluctuating forces between the two models is approximately 7:6. When the damping noise is equal to the intrinsic damping thermal noise, under the same pressure, the ratio of the gap spacing $h_{\mathrm{c}}$ constraints between the two models is approximately 7:6. In other words, the latter is about 1/7 lower than the former. As long as the base damping can be neglected, this result will not change with variations in experimental parameters.

\section{\label{sec:diffusiontime} Diffusion Time for Bi-plate Finite Open Systems}

Reference~\cite{Mao_2023} provides theoretical result for the diffusion time $\tau_{\mathrm{cuboid}}$ of square bi-plates:
\begin{equation}
	\begin{split}
		\tau_{x}
		=&\frac{b}{\sqrt{72\pi k_{\mathrm{B}}T/m}}\frac{1}{\varsigma\left(\sqrt{1+\varsigma^{2}}-\varsigma\right)}\\
		&\times \frac{2\arctan (1/\varsigma)+2\varsigma\ln\varsigma-\varsigma\ln(1+\varsigma^{2})}{2\varsigma-2\sqrt{1+\varsigma^{2}}+\ln\left(1+\sqrt{1+\varsigma^{2}}\right)-\ln\varsigma},
	\end{split}
\end{equation}
where $\varsigma=h/2b$, $a$ and $b$ are the side lengths of the square plates along the $x$ and $y$ axes, respectively. Similarly, $\tau_{y}$ can be calculated, and the total diffusion time is given by:
\begin{equation}
	\tau_{\mathrm{cuboid}}=\frac{a}{a+b}\tau_{x}+\frac{b}{a+b}\tau_{y}.
\end{equation}
Additionally, Ref.~\cite{Mao_2023} also provides diffusion time $\tau_{\mathrm{circle}}$ of circular bi-plate.

The bi-plate model has no special requirements on the shape of the plate. For example, it can be a ring, a triangle, or a more complex shape. However the diffusion time needs to be calculated through numerical simulation. However, in sensor cores, the gap boundary is different. For accurate diffusion time, numerical simulations are needed based on the specific geometric shape.

\bibliography{finiteopen.bib}

\providecommand{\noopsort}[1]{}\providecommand{\singleletter}[1]{#1}%
\begin{thebibliography}{91}%
\makeatletter
\providecommand \@ifxundefined [1]{%
 \@ifx{#1\undefined}
}%
\providecommand \@ifnum [1]{%
 \ifnum #1\expandafter \@firstoftwo
 \else \expandafter \@secondoftwo
 \fi
}%
\providecommand \@ifx [1]{%
 \ifx #1\expandafter \@firstoftwo
 \else \expandafter \@secondoftwo
 \fi
}%
\providecommand \natexlab [1]{#1}%
\providecommand \enquote  [1]{``#1''}%
\providecommand \bibnamefont  [1]{#1}%
\providecommand \bibfnamefont [1]{#1}%
\providecommand \citenamefont [1]{#1}%
\providecommand \href@noop [0]{\@secondoftwo}%
\providecommand \href [0]{\begingroup \@sanitize@url \@href}%
\providecommand \@href[1]{\@@startlink{#1}\@@href}%
\providecommand \@@href[1]{\endgroup#1\@@endlink}%
\providecommand \@sanitize@url [0]{\catcode `\\12\catcode `\$12\catcode
  `\&12\catcode `\#12\catcode `\^12\catcode `\_12\catcode `\%12\relax}%
\providecommand \@@startlink[1]{}%
\providecommand \@@endlink[0]{}%
\providecommand \url  [0]{\begingroup\@sanitize@url \@url }%
\providecommand \@url [1]{\endgroup\@href {#1}{\urlprefix }}%
\providecommand \urlprefix  [0]{URL }%
\providecommand \Eprint [0]{\href }%
\providecommand \doibase [0]{https://doi.org/}%
\providecommand \selectlanguage [0]{\@gobble}%
\providecommand \bibinfo  [0]{\@secondoftwo}%
\providecommand \bibfield  [0]{\@secondoftwo}%
\providecommand \translation [1]{[#1]}%
\providecommand \BibitemOpen [0]{}%
\providecommand \bibitemStop [0]{}%
\providecommand \bibitemNoStop [0]{.\EOS\space}%
\providecommand \EOS [0]{\spacefactor3000\relax}%
\providecommand \BibitemShut  [1]{\csname bibitem#1\endcsname}%
\let\auto@bib@innerbib\@empty
\bibitem [{\citenamefont {Dicke}(1964)}]{dicke1964experimental}%
  \BibitemOpen
  \bibfield  {author} {\bibinfo {author} {\bibfnamefont {R.~H.}\ \bibnamefont
  {Dicke}},\ }\bibfield  {title} {\bibinfo {title} {Experimental relativity:
  Relativity, groups, and topology},\ }in\ \href@noop {} {\emph {\bibinfo
  {booktitle} {Relativity, Groups and Topology}}}\ (\bibinfo  {publisher} {New
  York: Gordon and Breach},\ \bibinfo {year} {1964})\ pp.\ \bibinfo {pages}
  {165--313}\BibitemShut {NoStop}%
\bibitem [{\citenamefont {Will}(2014)}]{will2014confrontation}%
  \BibitemOpen
  \bibfield  {author} {\bibinfo {author} {\bibfnamefont {C.~M.}\ \bibnamefont
  {Will}},\ }\bibfield  {title} {\bibinfo {title} {The confrontation between
  general relativity and experiment},\ }\href
  {https://doi.org/10.12942/lrr-2014-4} {\bibfield  {journal} {\bibinfo
  {journal} {Living Rev. Relativ.}\ }\textbf {\bibinfo {volume} {17}},\
  \bibinfo {pages} {1} (\bibinfo {year} {2014})}\BibitemShut {NoStop}%
\bibitem [{\citenamefont {Will}(2018)}]{will2018theory}%
  \BibitemOpen
  \bibfield  {author} {\bibinfo {author} {\bibfnamefont {C.~M.}\ \bibnamefont
  {Will}},\ }\href@noop {} {\emph {\bibinfo {title} {Theory and Experiment in
  Gravitational Physics}}}\ (\bibinfo  {publisher} {Cambridge university
  press},\ \bibinfo {year} {2018})\BibitemShut {NoStop}%
\bibitem [{\citenamefont {Touboul}\ \emph {et~al.}(2002)\citenamefont
  {Touboul}, \citenamefont {Foulon}, \citenamefont {Lafargue},\ and\
  \citenamefont {Metris}}]{TOUBOUL2002433}%
  \BibitemOpen
  \bibfield  {author} {\bibinfo {author} {\bibfnamefont {P.}~\bibnamefont
  {Touboul}}, \bibinfo {author} {\bibfnamefont {B.}~\bibnamefont {Foulon}},
  \bibinfo {author} {\bibfnamefont {L.}~\bibnamefont {Lafargue}},\ and\
  \bibinfo {author} {\bibfnamefont {G.}~\bibnamefont {Metris}},\ }\bibfield
  {title} {\bibinfo {title} {The {MICROSCOPE} mission},\ }\href
  {https://doi.org/https://doi.org/10.1016/S0094-5765(01)00188-6} {\bibfield
  {journal} {\bibinfo  {journal} {Acta Astronaut.}\ }\textbf {\bibinfo {volume}
  {50}},\ \bibinfo {pages} {433} (\bibinfo {year} {2002})}\BibitemShut
  {NoStop}%
\bibitem [{\citenamefont {Touboul}\ \emph {et~al.}(2022)\citenamefont
  {Touboul}, \citenamefont {M\'etris}, \citenamefont {Rodrigues}, \citenamefont
  {Berg\'e}, \citenamefont {Robert}, \citenamefont {Baghi} \emph
  {et~al.}}]{PhysRevLett.129.121102}%
  \BibitemOpen
  \bibfield  {author} {\bibinfo {author} {\bibfnamefont {P.}~\bibnamefont
  {Touboul}}, \bibinfo {author} {\bibfnamefont {G.}~\bibnamefont {M\'etris}},
  \bibinfo {author} {\bibfnamefont {M.}~\bibnamefont {Rodrigues}}, \bibinfo
  {author} {\bibfnamefont {J.}~\bibnamefont {Berg\'e}}, \bibinfo {author}
  {\bibfnamefont {A.}~\bibnamefont {Robert}}, \bibinfo {author} {\bibfnamefont
  {Q.}~\bibnamefont {Baghi}}, \emph {et~al.} (\bibinfo {collaboration}
  {MICROSCOPE Collaboration}),\ }\bibfield  {title} {\bibinfo {title}
  {{MICROSCOPE} mission: Final results of the test of the equivalence
  principle},\ }\href {https://doi.org/10.1103/PhysRevLett.129.121102}
  {\bibfield  {journal} {\bibinfo  {journal} {Phys. Rev. Lett.}\ }\textbf
  {\bibinfo {volume} {129}},\ \bibinfo {pages} {121102} (\bibinfo {year}
  {2022})}\BibitemShut {NoStop}%
\bibitem [{\citenamefont {Touboul}\ \emph {et~al.}(2017)\citenamefont
  {Touboul}, \citenamefont {M\'etris}, \citenamefont {Rodrigues}, \citenamefont
  {Andr\'e}, \citenamefont {Baghi}, \citenamefont {Berg\'e} \emph
  {et~al.}}]{PhysRevLett.119.231101}%
  \BibitemOpen
  \bibfield  {author} {\bibinfo {author} {\bibfnamefont {P.}~\bibnamefont
  {Touboul}}, \bibinfo {author} {\bibfnamefont {G.}~\bibnamefont {M\'etris}},
  \bibinfo {author} {\bibfnamefont {M.}~\bibnamefont {Rodrigues}}, \bibinfo
  {author} {\bibfnamefont {Y.}~\bibnamefont {Andr\'e}}, \bibinfo {author}
  {\bibfnamefont {Q.}~\bibnamefont {Baghi}}, \bibinfo {author} {\bibfnamefont
  {J.}~\bibnamefont {Berg\'e}}, \emph {et~al.},\ }\bibfield  {title} {\bibinfo
  {title} {{MICROSCOPE} mission: First results of a space test of the
  equivalence principle},\ }\href
  {https://doi.org/10.1103/PhysRevLett.119.231101} {\bibfield  {journal}
  {\bibinfo  {journal} {Phys. Rev. Lett.}\ }\textbf {\bibinfo {volume} {119}},\
  \bibinfo {pages} {231101} (\bibinfo {year} {2017})}\BibitemShut {NoStop}%
\bibitem [{\citenamefont {Abbott}\ \emph {et~al.}(2016)\citenamefont {Abbott},
  \citenamefont {Abbott}, \citenamefont {Abbott}, \citenamefont {Abernathy},
  \citenamefont {Acernese}, \citenamefont {Ackley} \emph
  {et~al.}}]{PhysRevLett.116.061102}%
  \BibitemOpen
  \bibfield  {author} {\bibinfo {author} {\bibfnamefont {B.~P.}\ \bibnamefont
  {Abbott}}, \bibinfo {author} {\bibfnamefont {R.}~\bibnamefont {Abbott}},
  \bibinfo {author} {\bibfnamefont {T.~D.}\ \bibnamefont {Abbott}}, \bibinfo
  {author} {\bibfnamefont {M.~R.}\ \bibnamefont {Abernathy}}, \bibinfo {author}
  {\bibfnamefont {F.}~\bibnamefont {Acernese}}, \bibinfo {author}
  {\bibfnamefont {K.}~\bibnamefont {Ackley}}, \emph {et~al.} (\bibinfo
  {collaboration} {{LIGO} Scientific Collaboration and Virgo Collaboration}),\
  }\bibfield  {title} {\bibinfo {title} {Observation of gravitational waves
  from a binary black hole merger},\ }\href
  {https://doi.org/10.1103/PhysRevLett.116.061102} {\bibfield  {journal}
  {\bibinfo  {journal} {Phys. Rev. Lett.}\ }\textbf {\bibinfo {volume} {116}},\
  \bibinfo {pages} {061102} (\bibinfo {year} {2016})}\BibitemShut {NoStop}%
\bibitem [{\citenamefont {Buikema}\ \emph {et~al.}(2020)\citenamefont
  {Buikema}, \citenamefont {Cahillane}, \citenamefont {Mansell}, \citenamefont
  {Blair}, \citenamefont {Abbott}, \citenamefont {Adams} \emph
  {et~al.}}]{PhysRevD.102.062003}%
  \BibitemOpen
  \bibfield  {author} {\bibinfo {author} {\bibfnamefont {A.}~\bibnamefont
  {Buikema}}, \bibinfo {author} {\bibfnamefont {C.}~\bibnamefont {Cahillane}},
  \bibinfo {author} {\bibfnamefont {G.~L.}\ \bibnamefont {Mansell}}, \bibinfo
  {author} {\bibfnamefont {C.~D.}\ \bibnamefont {Blair}}, \bibinfo {author}
  {\bibfnamefont {R.}~\bibnamefont {Abbott}}, \bibinfo {author} {\bibfnamefont
  {C.}~\bibnamefont {Adams}}, \emph {et~al.},\ }\bibfield  {title} {\bibinfo
  {title} {Sensitivity and performance of the advanced {LIGO} detectors in the
  third observing run},\ }\href {https://doi.org/10.1103/PhysRevD.102.062003}
  {\bibfield  {journal} {\bibinfo  {journal} {Phys. Rev. D}\ }\textbf {\bibinfo
  {volume} {102}},\ \bibinfo {pages} {062003} (\bibinfo {year}
  {2020})}\BibitemShut {NoStop}%
\bibitem [{\citenamefont {Ganapathy}\ \emph {et~al.}(2023)\citenamefont
  {Ganapathy}, \citenamefont {Jia}, \citenamefont {Nakano}, \citenamefont {Xu},
  \citenamefont {Aritomi}, \citenamefont {Cullen} \emph
  {et~al.}}]{PhysRevX.13.041021}%
  \BibitemOpen
  \bibfield  {author} {\bibinfo {author} {\bibfnamefont {D.}~\bibnamefont
  {Ganapathy}}, \bibinfo {author} {\bibfnamefont {W.}~\bibnamefont {Jia}},
  \bibinfo {author} {\bibfnamefont {M.}~\bibnamefont {Nakano}}, \bibinfo
  {author} {\bibfnamefont {V.}~\bibnamefont {Xu}}, \bibinfo {author}
  {\bibfnamefont {N.}~\bibnamefont {Aritomi}}, \bibinfo {author} {\bibfnamefont
  {T.}~\bibnamefont {Cullen}}, \emph {et~al.} (\bibinfo {collaboration} {{LIGO}
  O4 Detector Collaboration}),\ }\bibfield  {title} {\bibinfo {title}
  {Broadband quantum enhancement of the {LIGO} detectors with
  frequency-dependent squeezing},\ }\href
  {https://doi.org/10.1103/PhysRevX.13.041021} {\bibfield  {journal} {\bibinfo
  {journal} {Phys. Rev. X}\ }\textbf {\bibinfo {volume} {13}},\ \bibinfo
  {pages} {041021} (\bibinfo {year} {2023})}\BibitemShut {NoStop}%
\bibitem [{\citenamefont {Armano}\ \emph {et~al.}(2016)\citenamefont {Armano},
  \citenamefont {Audley}, \citenamefont {Auger}, \citenamefont {Baird},
  \citenamefont {Bassan}, \citenamefont {Binetruy} \emph
  {et~al.}}]{PhysRevLett.116.231101}%
  \BibitemOpen
  \bibfield  {author} {\bibinfo {author} {\bibfnamefont {M.}~\bibnamefont
  {Armano}}, \bibinfo {author} {\bibfnamefont {H.}~\bibnamefont {Audley}},
  \bibinfo {author} {\bibfnamefont {G.}~\bibnamefont {Auger}}, \bibinfo
  {author} {\bibfnamefont {J.~T.}\ \bibnamefont {Baird}}, \bibinfo {author}
  {\bibfnamefont {M.}~\bibnamefont {Bassan}}, \bibinfo {author} {\bibfnamefont
  {P.}~\bibnamefont {Binetruy}}, \emph {et~al.},\ }\bibfield  {title} {\bibinfo
  {title} {Sub-femto-$g$ free fall for space-based gravitational wave
  observatories: {LISA Pathfinder} results},\ }\href
  {https://doi.org/10.1103/PhysRevLett.116.231101} {\bibfield  {journal}
  {\bibinfo  {journal} {Phys. Rev. Lett.}\ }\textbf {\bibinfo {volume} {116}},\
  \bibinfo {pages} {231101} (\bibinfo {year} {2016})}\BibitemShut {NoStop}%
\bibitem [{\citenamefont {Armano}\ \emph {et~al.}(2018)\citenamefont {Armano},
  \citenamefont {Audley}, \citenamefont {Baird}, \citenamefont {Binetruy},
  \citenamefont {Born}, \citenamefont {Bortoluzzi} \emph
  {et~al.}}]{PhysRevLett.120.061101}%
  \BibitemOpen
  \bibfield  {author} {\bibinfo {author} {\bibfnamefont {M.}~\bibnamefont
  {Armano}}, \bibinfo {author} {\bibfnamefont {H.}~\bibnamefont {Audley}},
  \bibinfo {author} {\bibfnamefont {J.}~\bibnamefont {Baird}}, \bibinfo
  {author} {\bibfnamefont {P.}~\bibnamefont {Binetruy}}, \bibinfo {author}
  {\bibfnamefont {M.}~\bibnamefont {Born}}, \bibinfo {author} {\bibfnamefont
  {D.}~\bibnamefont {Bortoluzzi}}, \emph {et~al.},\ }\bibfield  {title}
  {\bibinfo {title} {Beyond the required {LISA} free-fall performance: New
  {LISA Pathfinder} results down to $20\text{ }\text{
  }\ensuremath{\mu}\mathrm{Hz}$},\ }\href
  {https://doi.org/10.1103/PhysRevLett.120.061101} {\bibfield  {journal}
  {\bibinfo  {journal} {Phys. Rev. Lett.}\ }\textbf {\bibinfo {volume} {120}},\
  \bibinfo {pages} {061101} (\bibinfo {year} {2018})}\BibitemShut {NoStop}%
\bibitem [{\citenamefont {Luo}\ \emph {et~al.}(2016)\citenamefont {Luo},
  \citenamefont {Chen}, \citenamefont {Duan}, \citenamefont {Gong},
  \citenamefont {Hu}, \citenamefont {Ji} \emph {et~al.}}]{Luo_2016}%
  \BibitemOpen
  \bibfield  {author} {\bibinfo {author} {\bibfnamefont {J.}~\bibnamefont
  {Luo}}, \bibinfo {author} {\bibfnamefont {L.-S.}\ \bibnamefont {Chen}},
  \bibinfo {author} {\bibfnamefont {H.-Z.}\ \bibnamefont {Duan}}, \bibinfo
  {author} {\bibfnamefont {Y.-G.}\ \bibnamefont {Gong}}, \bibinfo {author}
  {\bibfnamefont {S.}~\bibnamefont {Hu}}, \bibinfo {author} {\bibfnamefont
  {J.}~\bibnamefont {Ji}}, \emph {et~al.},\ }\bibfield  {title} {\bibinfo
  {title} {{TianQin}: a space-borne gravitational wave detector},\ }\href
  {https://doi.org/10.1088/0264-9381/33/3/035010} {\bibfield  {journal}
  {\bibinfo  {journal} {Classical Quantum Gravity}\ }\textbf {\bibinfo {volume}
  {33}},\ \bibinfo {pages} {035010} (\bibinfo {year} {2016})}\BibitemShut
  {NoStop}%
\bibitem [{\citenamefont {Hu}\ and\ \citenamefont
  {Wu}(2017)}]{10.1093/nsr/nwx116}%
  \BibitemOpen
  \bibfield  {author} {\bibinfo {author} {\bibfnamefont {W.-R.}\ \bibnamefont
  {Hu}}\ and\ \bibinfo {author} {\bibfnamefont {Y.-L.}\ \bibnamefont {Wu}},\
  }\bibfield  {title} {\bibinfo {title} {The {Taiji} program in space for
  gravitational wave physics and the nature of gravity},\ }\href
  {https://doi.org/10.1093/nsr/nwx116} {\bibfield  {journal} {\bibinfo
  {journal} {Natl. Sci. Rev.}\ }\textbf {\bibinfo {volume} {4}},\ \bibinfo
  {pages} {685} (\bibinfo {year} {2017})}\BibitemShut {NoStop}%
\bibitem [{\citenamefont {Naeimi}\ and\ \citenamefont
  {Flury}(2017)}]{naeimi2017global}%
  \BibitemOpen
  \bibfield  {author} {\bibinfo {author} {\bibfnamefont {M.}~\bibnamefont
  {Naeimi}}\ and\ \bibinfo {author} {\bibfnamefont {J.}~\bibnamefont {Flury}},\
  }\href@noop {} {\emph {\bibinfo {title} {Global Gravity Field Modeling from
  Satellite-to-Satellite Tracking Data}}}\ (\bibinfo  {publisher} {Springer},\
  \bibinfo {year} {2017})\BibitemShut {NoStop}%
\bibitem [{\citenamefont {Kornfeld}\ \emph {et~al.}(2019)\citenamefont
  {Kornfeld}, \citenamefont {Arnold}, \citenamefont {Gross}, \citenamefont
  {Dahya}, \citenamefont {Klipstein}, \citenamefont {Gath},\ and\ \citenamefont
  {Bettadpur}}]{10.2514/1.A34326}%
  \BibitemOpen
  \bibfield  {author} {\bibinfo {author} {\bibfnamefont {R.~P.}\ \bibnamefont
  {Kornfeld}}, \bibinfo {author} {\bibfnamefont {B.~W.}\ \bibnamefont
  {Arnold}}, \bibinfo {author} {\bibfnamefont {M.~A.}\ \bibnamefont {Gross}},
  \bibinfo {author} {\bibfnamefont {N.~T.}\ \bibnamefont {Dahya}}, \bibinfo
  {author} {\bibfnamefont {W.~M.}\ \bibnamefont {Klipstein}}, \bibinfo {author}
  {\bibfnamefont {P.~F.}\ \bibnamefont {Gath}},\ and\ \bibinfo {author}
  {\bibfnamefont {S.}~\bibnamefont {Bettadpur}},\ }\bibfield  {title} {\bibinfo
  {title} {{GRACE-FO}: The gravity recovery and climate experiment follow-on
  mission},\ }\href {https://doi.org/10.2514/1.A34326} {\bibfield  {journal}
  {\bibinfo  {journal} {J. Spacecr. Rockets}\ }\textbf {\bibinfo {volume}
  {56}},\ \bibinfo {pages} {931} (\bibinfo {year} {2019})}\BibitemShut
  {NoStop}%
\bibitem [{\citenamefont {Touboul}\ \emph {et~al.}(2012)\citenamefont
  {Touboul}, \citenamefont {Foulon}, \citenamefont {Christophe},\ and\
  \citenamefont {Marque}}]{touboul2012champ}%
  \BibitemOpen
  \bibfield  {author} {\bibinfo {author} {\bibfnamefont {P.}~\bibnamefont
  {Touboul}}, \bibinfo {author} {\bibfnamefont {B.}~\bibnamefont {Foulon}},
  \bibinfo {author} {\bibfnamefont {B.}~\bibnamefont {Christophe}},\ and\
  \bibinfo {author} {\bibfnamefont {J.}~\bibnamefont {Marque}},\ }\bibfield
  {title} {\bibinfo {title} {{CHAMP, GRACE, GOCE instruments and beyond}},\
  }in\ \href@noop {} {\emph {\bibinfo {booktitle} {Geodesy for Planet Earth:
  Proceedings of the 2009 IAG Symposium, Buenos Aires, Argentina, 31 August
  31-4 September 2009}}}\ (\bibinfo {organization} {Springer},\ \bibinfo {year}
  {2012})\ pp.\ \bibinfo {pages} {215--221}\BibitemShut {NoStop}%
\bibitem [{\citenamefont {Adelberger}\ \emph {et~al.}(2007)\citenamefont
  {Adelberger}, \citenamefont {Heckel}, \citenamefont {Hoedl}, \citenamefont
  {Hoyle}, \citenamefont {Kapner},\ and\ \citenamefont
  {Upadhye}}]{PhysRevLett.98.131104}%
  \BibitemOpen
  \bibfield  {author} {\bibinfo {author} {\bibfnamefont {E.~G.}\ \bibnamefont
  {Adelberger}}, \bibinfo {author} {\bibfnamefont {B.~R.}\ \bibnamefont
  {Heckel}}, \bibinfo {author} {\bibfnamefont {S.}~\bibnamefont {Hoedl}},
  \bibinfo {author} {\bibfnamefont {C.~D.}\ \bibnamefont {Hoyle}}, \bibinfo
  {author} {\bibfnamefont {D.~J.}\ \bibnamefont {Kapner}},\ and\ \bibinfo
  {author} {\bibfnamefont {A.}~\bibnamefont {Upadhye}},\ }\bibfield  {title}
  {\bibinfo {title} {Particle-physics implications of a recent test of the
  gravitational inverse-square law},\ }\href
  {https://doi.org/10.1103/PhysRevLett.98.131104} {\bibfield  {journal}
  {\bibinfo  {journal} {Phys. Rev. Lett.}\ }\textbf {\bibinfo {volume} {98}},\
  \bibinfo {pages} {131104} (\bibinfo {year} {2007})}\BibitemShut {NoStop}%
\bibitem [{\citenamefont {Yang}\ \emph {et~al.}(2012)\citenamefont {Yang},
  \citenamefont {Zhan}, \citenamefont {Wang}, \citenamefont {Shao},
  \citenamefont {Tu}, \citenamefont {Tan},\ and\ \citenamefont
  {Luo}}]{PhysRevLett.108.081101}%
  \BibitemOpen
  \bibfield  {author} {\bibinfo {author} {\bibfnamefont {S.-Q.}\ \bibnamefont
  {Yang}}, \bibinfo {author} {\bibfnamefont {B.-F.}\ \bibnamefont {Zhan}},
  \bibinfo {author} {\bibfnamefont {Q.-L.}\ \bibnamefont {Wang}}, \bibinfo
  {author} {\bibfnamefont {C.-G.}\ \bibnamefont {Shao}}, \bibinfo {author}
  {\bibfnamefont {L.-C.}\ \bibnamefont {Tu}}, \bibinfo {author} {\bibfnamefont
  {W.-H.}\ \bibnamefont {Tan}},\ and\ \bibinfo {author} {\bibfnamefont
  {J.}~\bibnamefont {Luo}},\ }\bibfield  {title} {\bibinfo {title} {Test of the
  gravitational inverse square law at millimeter ranges},\ }\href
  {https://doi.org/10.1103/PhysRevLett.108.081101} {\bibfield  {journal}
  {\bibinfo  {journal} {Phys. Rev. Lett.}\ }\textbf {\bibinfo {volume} {108}},\
  \bibinfo {pages} {081101} (\bibinfo {year} {2012})}\BibitemShut {NoStop}%
\bibitem [{\citenamefont {Tan}\ \emph {et~al.}(2016)\citenamefont {Tan},
  \citenamefont {Yang}, \citenamefont {Shao}, \citenamefont {Li}, \citenamefont
  {Du}, \citenamefont {Zhan}, \citenamefont {Wang}, \citenamefont {Luo},
  \citenamefont {Tu},\ and\ \citenamefont {Luo}}]{PhysRevLett.116.131101}%
  \BibitemOpen
  \bibfield  {author} {\bibinfo {author} {\bibfnamefont {W.-H.}\ \bibnamefont
  {Tan}}, \bibinfo {author} {\bibfnamefont {S.-Q.}\ \bibnamefont {Yang}},
  \bibinfo {author} {\bibfnamefont {C.-G.}\ \bibnamefont {Shao}}, \bibinfo
  {author} {\bibfnamefont {J.}~\bibnamefont {Li}}, \bibinfo {author}
  {\bibfnamefont {A.-B.}\ \bibnamefont {Du}}, \bibinfo {author} {\bibfnamefont
  {B.-F.}\ \bibnamefont {Zhan}}, \bibinfo {author} {\bibfnamefont {Q.-L.}\
  \bibnamefont {Wang}}, \bibinfo {author} {\bibfnamefont {P.-S.}\ \bibnamefont
  {Luo}}, \bibinfo {author} {\bibfnamefont {L.-C.}\ \bibnamefont {Tu}},\ and\
  \bibinfo {author} {\bibfnamefont {J.}~\bibnamefont {Luo}},\ }\bibfield
  {title} {\bibinfo {title} {New test of the gravitational inverse-square law
  at the submillimeter range with dual modulation and compensation},\ }\href
  {https://doi.org/10.1103/PhysRevLett.116.131101} {\bibfield  {journal}
  {\bibinfo  {journal} {Phys. Rev. Lett.}\ }\textbf {\bibinfo {volume} {116}},\
  \bibinfo {pages} {131101} (\bibinfo {year} {2016})}\BibitemShut {NoStop}%
\bibitem [{\citenamefont {Tan}\ \emph {et~al.}(2020)\citenamefont {Tan},
  \citenamefont {Du}, \citenamefont {Dong}, \citenamefont {Yang}, \citenamefont
  {Shao}, \citenamefont {Guan}, \citenamefont {Wang}, \citenamefont {Zhan},
  \citenamefont {Luo}, \citenamefont {Tu},\ and\ \citenamefont
  {Luo}}]{PhysRevLett.124.051301}%
  \BibitemOpen
  \bibfield  {author} {\bibinfo {author} {\bibfnamefont {W.-H.}\ \bibnamefont
  {Tan}}, \bibinfo {author} {\bibfnamefont {A.-B.}\ \bibnamefont {Du}},
  \bibinfo {author} {\bibfnamefont {W.-C.}\ \bibnamefont {Dong}}, \bibinfo
  {author} {\bibfnamefont {S.-Q.}\ \bibnamefont {Yang}}, \bibinfo {author}
  {\bibfnamefont {C.-G.}\ \bibnamefont {Shao}}, \bibinfo {author}
  {\bibfnamefont {S.-G.}\ \bibnamefont {Guan}}, \bibinfo {author}
  {\bibfnamefont {Q.-L.}\ \bibnamefont {Wang}}, \bibinfo {author}
  {\bibfnamefont {B.-F.}\ \bibnamefont {Zhan}}, \bibinfo {author}
  {\bibfnamefont {P.-S.}\ \bibnamefont {Luo}}, \bibinfo {author} {\bibfnamefont
  {L.-C.}\ \bibnamefont {Tu}},\ and\ \bibinfo {author} {\bibfnamefont
  {J.}~\bibnamefont {Luo}},\ }\bibfield  {title} {\bibinfo {title} {Improvement
  for testing the gravitational inverse-square law at the submillimeter
  range},\ }\href {https://doi.org/10.1103/PhysRevLett.124.051301} {\bibfield
  {journal} {\bibinfo  {journal} {Phys. Rev. Lett.}\ }\textbf {\bibinfo
  {volume} {124}},\ \bibinfo {pages} {051301} (\bibinfo {year}
  {2020})}\BibitemShut {NoStop}%
\bibitem [{\citenamefont {Shao}\ \emph {et~al.}(2016)\citenamefont {Shao},
  \citenamefont {Tan}, \citenamefont {Tan}, \citenamefont {Yang}, \citenamefont
  {Luo}, \citenamefont {Tobar}, \citenamefont {Bailey}, \citenamefont {Long},
  \citenamefont {Weisman}, \citenamefont {Xu},\ and\ \citenamefont
  {Kosteleck\'y}}]{PhysRevLett.117.071102}%
  \BibitemOpen
  \bibfield  {author} {\bibinfo {author} {\bibfnamefont {C.-G.}\ \bibnamefont
  {Shao}}, \bibinfo {author} {\bibfnamefont {Y.-J.}\ \bibnamefont {Tan}},
  \bibinfo {author} {\bibfnamefont {W.-H.}\ \bibnamefont {Tan}}, \bibinfo
  {author} {\bibfnamefont {S.-Q.}\ \bibnamefont {Yang}}, \bibinfo {author}
  {\bibfnamefont {J.}~\bibnamefont {Luo}}, \bibinfo {author} {\bibfnamefont
  {M.~E.}\ \bibnamefont {Tobar}}, \bibinfo {author} {\bibfnamefont {Q.~G.}\
  \bibnamefont {Bailey}}, \bibinfo {author} {\bibfnamefont {J.~C.}\
  \bibnamefont {Long}}, \bibinfo {author} {\bibfnamefont {E.}~\bibnamefont
  {Weisman}}, \bibinfo {author} {\bibfnamefont {R.}~\bibnamefont {Xu}},\ and\
  \bibinfo {author} {\bibfnamefont {V.~A.}\ \bibnamefont {Kosteleck\'y}},\
  }\bibfield  {title} {\bibinfo {title} {Combined search for {Lorentz}
  violation in short-range gravity},\ }\href
  {https://doi.org/10.1103/PhysRevLett.117.071102} {\bibfield  {journal}
  {\bibinfo  {journal} {Phys. Rev. Lett.}\ }\textbf {\bibinfo {volume} {117}},\
  \bibinfo {pages} {071102} (\bibinfo {year} {2016})}\BibitemShut {NoStop}%
\bibitem [{\citenamefont {Shao}\ \emph {et~al.}(2019)\citenamefont {Shao},
  \citenamefont {Chen}, \citenamefont {Tan}, \citenamefont {Yang},
  \citenamefont {Luo}, \citenamefont {Tobar}, \citenamefont {Long},
  \citenamefont {Weisman},\ and\ \citenamefont
  {Kosteleck\'y}}]{PhysRevLett.122.011102}%
  \BibitemOpen
  \bibfield  {author} {\bibinfo {author} {\bibfnamefont {C.-G.}\ \bibnamefont
  {Shao}}, \bibinfo {author} {\bibfnamefont {Y.-F.}\ \bibnamefont {Chen}},
  \bibinfo {author} {\bibfnamefont {Y.-J.}\ \bibnamefont {Tan}}, \bibinfo
  {author} {\bibfnamefont {S.-Q.}\ \bibnamefont {Yang}}, \bibinfo {author}
  {\bibfnamefont {J.}~\bibnamefont {Luo}}, \bibinfo {author} {\bibfnamefont
  {M.~E.}\ \bibnamefont {Tobar}}, \bibinfo {author} {\bibfnamefont {J.~C.}\
  \bibnamefont {Long}}, \bibinfo {author} {\bibfnamefont {E.}~\bibnamefont
  {Weisman}},\ and\ \bibinfo {author} {\bibfnamefont {V.~A.}\ \bibnamefont
  {Kosteleck\'y}},\ }\bibfield  {title} {\bibinfo {title} {Combined search for
  a {Lorentz}-violating force in short-range gravity varying as the inverse
  sixth power of distance},\ }\href
  {https://doi.org/10.1103/PhysRevLett.122.011102} {\bibfield  {journal}
  {\bibinfo  {journal} {Phys. Rev. Lett.}\ }\textbf {\bibinfo {volume} {122}},\
  \bibinfo {pages} {011102} (\bibinfo {year} {2019})}\BibitemShut {NoStop}%
\bibitem [{\citenamefont {Ke}\ \emph {et~al.}(2021)\citenamefont {Ke},
  \citenamefont {Luo}, \citenamefont {Shao}, \citenamefont {Tan}, \citenamefont
  {Tan},\ and\ \citenamefont {Yang}}]{PhysRevLett.126.211101}%
  \BibitemOpen
  \bibfield  {author} {\bibinfo {author} {\bibfnamefont {J.}~\bibnamefont
  {Ke}}, \bibinfo {author} {\bibfnamefont {J.}~\bibnamefont {Luo}}, \bibinfo
  {author} {\bibfnamefont {C.-G.}\ \bibnamefont {Shao}}, \bibinfo {author}
  {\bibfnamefont {Y.-J.}\ \bibnamefont {Tan}}, \bibinfo {author} {\bibfnamefont
  {W.-H.}\ \bibnamefont {Tan}},\ and\ \bibinfo {author} {\bibfnamefont {S.-Q.}\
  \bibnamefont {Yang}},\ }\bibfield  {title} {\bibinfo {title} {Combined test
  of the gravitational inverse-square law at the centimeter range},\ }\href
  {https://doi.org/10.1103/PhysRevLett.126.211101} {\bibfield  {journal}
  {\bibinfo  {journal} {Phys. Rev. Lett.}\ }\textbf {\bibinfo {volume} {126}},\
  \bibinfo {pages} {211101} (\bibinfo {year} {2021})}\BibitemShut {NoStop}%
\bibitem [{\citenamefont {Li}\ \emph {et~al.}(2018)\citenamefont {Li} \emph
  {et~al.}}]{li2018measurements}%
  \BibitemOpen
  \bibfield  {author} {\bibinfo {author} {\bibfnamefont {Q.}~\bibnamefont {Li}}
  \emph {et~al.},\ }\bibfield  {title} {\bibinfo {title} {Measurements of the
  gravitational constant using two independent methods},\ }\href
  {https://doi.org/10.1038/s41586-018-0431-5} {\bibfield  {journal} {\bibinfo
  {journal} {Nature}\ }\textbf {\bibinfo {volume} {560}},\ \bibinfo {pages}
  {582} (\bibinfo {year} {2018})}\BibitemShut {NoStop}%
\bibitem [{\citenamefont {Tiesinga}\ \emph {et~al.}(2021)\citenamefont
  {Tiesinga}, \citenamefont {Mohr}, \citenamefont {Newell},\ and\ \citenamefont
  {Taylor}}]{RevModPhys.93.025010}%
  \BibitemOpen
  \bibfield  {author} {\bibinfo {author} {\bibfnamefont {E.}~\bibnamefont
  {Tiesinga}}, \bibinfo {author} {\bibfnamefont {P.~J.}\ \bibnamefont {Mohr}},
  \bibinfo {author} {\bibfnamefont {D.~B.}\ \bibnamefont {Newell}},\ and\
  \bibinfo {author} {\bibfnamefont {B.~N.}\ \bibnamefont {Taylor}},\ }\bibfield
   {title} {\bibinfo {title} {{CODATA} recommended values of the fundamental
  physical constants: 2018},\ }\href
  {https://doi.org/10.1103/RevModPhys.93.025010} {\bibfield  {journal}
  {\bibinfo  {journal} {Rev. Mod. Phys.}\ }\textbf {\bibinfo {volume} {93}},\
  \bibinfo {pages} {025010} (\bibinfo {year} {2021})}\BibitemShut {NoStop}%
\bibitem [{\citenamefont {Rosi}\ \emph {et~al.}(2014)\citenamefont {Rosi},
  \citenamefont {Sorrentino}, \citenamefont {Cacciapuoti}, \citenamefont
  {Prevedelli},\ and\ \citenamefont {Tino}}]{rosi2014precision}%
  \BibitemOpen
  \bibfield  {author} {\bibinfo {author} {\bibfnamefont {G.}~\bibnamefont
  {Rosi}}, \bibinfo {author} {\bibfnamefont {F.}~\bibnamefont {Sorrentino}},
  \bibinfo {author} {\bibfnamefont {L.}~\bibnamefont {Cacciapuoti}}, \bibinfo
  {author} {\bibfnamefont {M.}~\bibnamefont {Prevedelli}},\ and\ \bibinfo
  {author} {\bibfnamefont {G.}~\bibnamefont {Tino}},\ }\bibfield  {title}
  {\bibinfo {title} {Precision measurement of the {Newtonian} gravitational
  constant using cold atoms},\ }\href {https://doi.org/10.1038/nature13433}
  {\bibfield  {journal} {\bibinfo  {journal} {Nature}\ }\textbf {\bibinfo
  {volume} {510}},\ \bibinfo {pages} {518} (\bibinfo {year}
  {2014})}\BibitemShut {NoStop}%
\bibitem [{\citenamefont {Touboul}\ \emph {et~al.}(1999)\citenamefont
  {Touboul}, \citenamefont {Willemenot}, \citenamefont {Foulon},\ and\
  \citenamefont {Josselin}}]{touboul1999accelerometers}%
  \BibitemOpen
  \bibfield  {author} {\bibinfo {author} {\bibfnamefont {P.}~\bibnamefont
  {Touboul}}, \bibinfo {author} {\bibfnamefont {E.}~\bibnamefont {Willemenot}},
  \bibinfo {author} {\bibfnamefont {B.}~\bibnamefont {Foulon}},\ and\ \bibinfo
  {author} {\bibfnamefont {V.}~\bibnamefont {Josselin}},\ }\bibfield  {title}
  {\bibinfo {title} {Accelerometers for {CHAMP, GRACE and GOCE} space missions:
  synergy and evolution},\ }\href@noop {} {\bibfield  {journal} {\bibinfo
  {journal} {Boll. Geof. Teor. Appl}\ }\textbf {\bibinfo {volume} {40}},\
  \bibinfo {pages} {321} (\bibinfo {year} {1999})}\BibitemShut {NoStop}%
\bibitem [{\citenamefont {Abramovici}\ \emph {et~al.}(1992)\citenamefont
  {Abramovici}, \citenamefont {Althouse}, \citenamefont {Drever}, \citenamefont
  {Gürsel}, \citenamefont {Kawamura}, \citenamefont {Raab}, \citenamefont
  {Shoemaker}, \citenamefont {Sievers}, \citenamefont {Spero}, \citenamefont
  {Thorne}, \citenamefont {Vogt}, \citenamefont {Weiss}, \citenamefont
  {Whitcomb},\ and\ \citenamefont {Zucker}}]{doi:10.1126/science.256.5055.325}%
  \BibitemOpen
  \bibfield  {author} {\bibinfo {author} {\bibfnamefont {A.}~\bibnamefont
  {Abramovici}}, \bibinfo {author} {\bibfnamefont {W.~E.}\ \bibnamefont
  {Althouse}}, \bibinfo {author} {\bibfnamefont {R.~W.~P.}\ \bibnamefont
  {Drever}}, \bibinfo {author} {\bibfnamefont {Y.}~\bibnamefont {Gürsel}},
  \bibinfo {author} {\bibfnamefont {S.}~\bibnamefont {Kawamura}}, \bibinfo
  {author} {\bibfnamefont {F.~J.}\ \bibnamefont {Raab}}, \bibinfo {author}
  {\bibfnamefont {D.}~\bibnamefont {Shoemaker}}, \bibinfo {author}
  {\bibfnamefont {L.}~\bibnamefont {Sievers}}, \bibinfo {author} {\bibfnamefont
  {R.~E.}\ \bibnamefont {Spero}}, \bibinfo {author} {\bibfnamefont {K.~S.}\
  \bibnamefont {Thorne}}, \bibinfo {author} {\bibfnamefont {R.~E.}\
  \bibnamefont {Vogt}}, \bibinfo {author} {\bibfnamefont {R.}~\bibnamefont
  {Weiss}}, \bibinfo {author} {\bibfnamefont {S.~E.}\ \bibnamefont
  {Whitcomb}},\ and\ \bibinfo {author} {\bibfnamefont {M.~E.}\ \bibnamefont
  {Zucker}},\ }\bibfield  {title} {\bibinfo {title} {{LIGO}: The laser
  interferometer gravitational-wave observatory},\ }\href
  {https://doi.org/10.1126/science.256.5055.325} {\bibfield  {journal}
  {\bibinfo  {journal} {Science}\ }\textbf {\bibinfo {volume} {256}},\ \bibinfo
  {pages} {325} (\bibinfo {year} {1992})}\BibitemShut {NoStop}%
\bibitem [{\citenamefont {Paik}\ \emph {et~al.}(2020)\citenamefont {Paik},
  \citenamefont {Vol~Moody},\ and\ \citenamefont
  {Norton}}]{doi:10.1142/S0218271819400017}%
  \BibitemOpen
  \bibfield  {author} {\bibinfo {author} {\bibfnamefont {H.~J.}\ \bibnamefont
  {Paik}}, \bibinfo {author} {\bibfnamefont {M.}~\bibnamefont {Vol~Moody}},\
  and\ \bibinfo {author} {\bibfnamefont {R.~S.}\ \bibnamefont {Norton}},\
  }\bibfield  {title} {\bibinfo {title} {{SOGRO} --- terrestrial full-tensor
  detector for mid-frequency gravitational waves},\ }\href
  {https://doi.org/10.1142/S0218271819400017} {\bibfield  {journal} {\bibinfo
  {journal} {Int. J. Mod. Phys. D}\ }\textbf {\bibinfo {volume} {29}},\
  \bibinfo {pages} {1940001} (\bibinfo {year} {2020})}\BibitemShut {NoStop}%
\bibitem [{\citenamefont {Schumaker}(2003)}]{BonnyLSchumaker_2003}%
  \BibitemOpen
  \bibfield  {author} {\bibinfo {author} {\bibfnamefont {B.~L.}\ \bibnamefont
  {Schumaker}},\ }\bibfield  {title} {\bibinfo {title} {Disturbance reduction
  requirements for {LISA}},\ }\href
  {https://doi.org/10.1088/0264-9381/20/10/327} {\bibfield  {journal} {\bibinfo
   {journal} {Classical Quantum Gravity}\ }\textbf {\bibinfo {volume} {20}},\
  \bibinfo {pages} {S239} (\bibinfo {year} {2003})}\BibitemShut {NoStop}%
\bibitem [{\citenamefont {Cavalleri}\ \emph {et~al.}(2009)\citenamefont
  {Cavalleri}, \citenamefont {Ciani}, \citenamefont {Dolesi}, \citenamefont
  {Heptonstall}, \citenamefont {Hueller}, \citenamefont {Nicolodi},
  \citenamefont {Rowan}, \citenamefont {Tombolato}, \citenamefont {Vitale},
  \citenamefont {Wass},\ and\ \citenamefont {Weber}}]{PhysRevLett.103.140601}%
  \BibitemOpen
  \bibfield  {author} {\bibinfo {author} {\bibfnamefont {A.}~\bibnamefont
  {Cavalleri}}, \bibinfo {author} {\bibfnamefont {G.}~\bibnamefont {Ciani}},
  \bibinfo {author} {\bibfnamefont {R.}~\bibnamefont {Dolesi}}, \bibinfo
  {author} {\bibfnamefont {A.}~\bibnamefont {Heptonstall}}, \bibinfo {author}
  {\bibfnamefont {M.}~\bibnamefont {Hueller}}, \bibinfo {author} {\bibfnamefont
  {D.}~\bibnamefont {Nicolodi}}, \bibinfo {author} {\bibfnamefont
  {S.}~\bibnamefont {Rowan}}, \bibinfo {author} {\bibfnamefont
  {D.}~\bibnamefont {Tombolato}}, \bibinfo {author} {\bibfnamefont
  {S.}~\bibnamefont {Vitale}}, \bibinfo {author} {\bibfnamefont {P.~J.}\
  \bibnamefont {Wass}},\ and\ \bibinfo {author} {\bibfnamefont {W.~J.}\
  \bibnamefont {Weber}},\ }\bibfield  {title} {\bibinfo {title} {Increased
  {Brownian} force noise from molecular impacts in a constrained volume},\
  }\href {https://doi.org/10.1103/PhysRevLett.103.140601} {\bibfield  {journal}
  {\bibinfo  {journal} {Phys. Rev. Lett.}\ }\textbf {\bibinfo {volume} {103}},\
  \bibinfo {pages} {140601} (\bibinfo {year} {2009})}\BibitemShut {NoStop}%
\bibitem [{\citenamefont {Dolesi}\ \emph {et~al.}(2011)\citenamefont {Dolesi},
  \citenamefont {Hueller}, \citenamefont {Nicolodi}, \citenamefont {Tombolato},
  \citenamefont {Vitale}, \citenamefont {Wass}, \citenamefont {Weber},
  \citenamefont {Evans}, \citenamefont {Fritschel}, \citenamefont {Weiss},
  \citenamefont {Gundlach}, \citenamefont {Hagedorn}, \citenamefont
  {Schlamminger}, \citenamefont {Ciani},\ and\ \citenamefont
  {Cavalleri}}]{PhysRevD.84.063007}%
  \BibitemOpen
  \bibfield  {author} {\bibinfo {author} {\bibfnamefont {R.}~\bibnamefont
  {Dolesi}}, \bibinfo {author} {\bibfnamefont {M.}~\bibnamefont {Hueller}},
  \bibinfo {author} {\bibfnamefont {D.}~\bibnamefont {Nicolodi}}, \bibinfo
  {author} {\bibfnamefont {D.}~\bibnamefont {Tombolato}}, \bibinfo {author}
  {\bibfnamefont {S.}~\bibnamefont {Vitale}}, \bibinfo {author} {\bibfnamefont
  {P.~J.}\ \bibnamefont {Wass}}, \bibinfo {author} {\bibfnamefont {W.~J.}\
  \bibnamefont {Weber}}, \bibinfo {author} {\bibfnamefont {M.}~\bibnamefont
  {Evans}}, \bibinfo {author} {\bibfnamefont {P.}~\bibnamefont {Fritschel}},
  \bibinfo {author} {\bibfnamefont {R.}~\bibnamefont {Weiss}}, \bibinfo
  {author} {\bibfnamefont {J.~H.}\ \bibnamefont {Gundlach}}, \bibinfo {author}
  {\bibfnamefont {C.~A.}\ \bibnamefont {Hagedorn}}, \bibinfo {author}
  {\bibfnamefont {S.}~\bibnamefont {Schlamminger}}, \bibinfo {author}
  {\bibfnamefont {G.}~\bibnamefont {Ciani}},\ and\ \bibinfo {author}
  {\bibfnamefont {A.}~\bibnamefont {Cavalleri}},\ }\bibfield  {title} {\bibinfo
  {title} {Brownian force noise from molecular collisions and the sensitivity
  of advanced gravitational wave observatories},\ }\href
  {https://doi.org/10.1103/PhysRevD.84.063007} {\bibfield  {journal} {\bibinfo
  {journal} {Phys. Rev. D}\ }\textbf {\bibinfo {volume} {84}},\ \bibinfo
  {pages} {063007} (\bibinfo {year} {2011})}\BibitemShut {NoStop}%
\bibitem [{\citenamefont {Carbone}\ \emph {et~al.}(2007)\citenamefont
  {Carbone}, \citenamefont {Cavalleri}, \citenamefont {Ciani}, \citenamefont
  {Dolesi}, \citenamefont {Hueller}, \citenamefont {Tombolato}, \citenamefont
  {Vitale},\ and\ \citenamefont {Weber}}]{PhysRevD.76.102003}%
  \BibitemOpen
  \bibfield  {author} {\bibinfo {author} {\bibfnamefont {L.}~\bibnamefont
  {Carbone}}, \bibinfo {author} {\bibfnamefont {A.}~\bibnamefont {Cavalleri}},
  \bibinfo {author} {\bibfnamefont {G.}~\bibnamefont {Ciani}}, \bibinfo
  {author} {\bibfnamefont {R.}~\bibnamefont {Dolesi}}, \bibinfo {author}
  {\bibfnamefont {M.}~\bibnamefont {Hueller}}, \bibinfo {author} {\bibfnamefont
  {D.}~\bibnamefont {Tombolato}}, \bibinfo {author} {\bibfnamefont
  {S.}~\bibnamefont {Vitale}},\ and\ \bibinfo {author} {\bibfnamefont {W.~J.}\
  \bibnamefont {Weber}},\ }\bibfield  {title} {\bibinfo {title} {Thermal
  gradient-induced forces on geodesic reference masses for {LISA}},\ }\href
  {https://doi.org/10.1103/PhysRevD.76.102003} {\bibfield  {journal} {\bibinfo
  {journal} {Phys. Rev. D}\ }\textbf {\bibinfo {volume} {76}},\ \bibinfo
  {pages} {102003} (\bibinfo {year} {2007})}\BibitemShut {NoStop}%
\bibitem [{\citenamefont {Xue}(2011)}]{DatongXue_2011}%
  \BibitemOpen
  \bibfield  {author} {\bibinfo {author} {\bibfnamefont {D.}~\bibnamefont
  {Xue}},\ }\bibfield  {title} {\bibinfo {title} {Noises sources with vacuum in
  electrostatically suspended accelerometer},\ }\href
  {https://kns.cnki.net/kcms2/article/abstract?v=fmMZJtqnKJYxLvJQUjm3ZJFUXEaoNogt8Ead8QQx_BqwG58I-cPT24cWALTo6Y6uRbdTica9Z3dPsXqI-5uuX_nskd60SjHH9HYgzVqMNfLKfHOBhR6WaMmWxzONRTUX&uniplatform=NZKPT&language=CHS}
  {\bibfield  {journal} {\bibinfo  {journal} {Chin. J. Vac. Sci. Technol.}\
  }\textbf {\bibinfo {volume} {31}},\ \bibinfo {pages} {633} (\bibinfo {year}
  {2011})}\BibitemShut {NoStop}%
\bibitem [{\citenamefont {Schlamminger}\ \emph {et~al.}(2010)\citenamefont
  {Schlamminger}, \citenamefont {Hagedorn},\ and\ \citenamefont
  {Gundlach}}]{PhysRevD.81.123008}%
  \BibitemOpen
  \bibfield  {author} {\bibinfo {author} {\bibfnamefont {S.}~\bibnamefont
  {Schlamminger}}, \bibinfo {author} {\bibfnamefont {C.~A.}\ \bibnamefont
  {Hagedorn}},\ and\ \bibinfo {author} {\bibfnamefont {J.~H.}\ \bibnamefont
  {Gundlach}},\ }\bibfield  {title} {\bibinfo {title} {Indirect evidence for
  {L\'evy} walks in squeeze film damping},\ }\href
  {https://doi.org/10.1103/PhysRevD.81.123008} {\bibfield  {journal} {\bibinfo
  {journal} {Phys. Rev. D}\ }\textbf {\bibinfo {volume} {81}},\ \bibinfo
  {pages} {123008} (\bibinfo {year} {2010})}\BibitemShut {NoStop}%
\bibitem [{\citenamefont {Cavalleri}\ \emph {et~al.}(2010)\citenamefont
  {Cavalleri}, \citenamefont {Ciani}, \citenamefont {Dolesi}, \citenamefont
  {Hueller}, \citenamefont {Nicolodi}, \citenamefont {Tombolato}, \citenamefont
  {Vitale}, \citenamefont {Wass},\ and\ \citenamefont
  {Weber}}]{CAVALLERI20103365}%
  \BibitemOpen
  \bibfield  {author} {\bibinfo {author} {\bibfnamefont {A.}~\bibnamefont
  {Cavalleri}}, \bibinfo {author} {\bibfnamefont {G.}~\bibnamefont {Ciani}},
  \bibinfo {author} {\bibfnamefont {R.}~\bibnamefont {Dolesi}}, \bibinfo
  {author} {\bibfnamefont {M.}~\bibnamefont {Hueller}}, \bibinfo {author}
  {\bibfnamefont {D.}~\bibnamefont {Nicolodi}}, \bibinfo {author}
  {\bibfnamefont {D.}~\bibnamefont {Tombolato}}, \bibinfo {author}
  {\bibfnamefont {S.}~\bibnamefont {Vitale}}, \bibinfo {author} {\bibfnamefont
  {P.}~\bibnamefont {Wass}},\ and\ \bibinfo {author} {\bibfnamefont
  {W.}~\bibnamefont {Weber}},\ }\bibfield  {title} {\bibinfo {title} {Gas
  damping force noise on a macroscopic test body in an infinite gas
  reservoir},\ }\href
  {https://doi.org/https://doi.org/10.1016/j.physleta.2010.06.041} {\bibfield
  {journal} {\bibinfo  {journal} {Phys. Lett. A}\ }\textbf {\bibinfo {volume}
  {374}},\ \bibinfo {pages} {3365} (\bibinfo {year} {2010})}\BibitemShut
  {NoStop}%
\bibitem [{\citenamefont {Ke}\ \emph {et~al.}(2020)\citenamefont {Ke},
  \citenamefont {Luo}, \citenamefont {Tan},\ and\ \citenamefont
  {Shao}}]{Ke_2020}%
  \BibitemOpen
  \bibfield  {author} {\bibinfo {author} {\bibfnamefont {J.}~\bibnamefont
  {Ke}}, \bibinfo {author} {\bibfnamefont {J.}~\bibnamefont {Luo}}, \bibinfo
  {author} {\bibfnamefont {Y.-J.}\ \bibnamefont {Tan}},\ and\ \bibinfo {author}
  {\bibfnamefont {C.-G.}\ \bibnamefont {Shao}},\ }\bibfield  {title} {\bibinfo
  {title} {Influence of the residual gas damping noise in the test of the
  gravitational inverse-square law},\ }\href
  {https://doi.org/10.1088/1361-6382/abb076} {\bibfield  {journal} {\bibinfo
  {journal} {Classical Quantum Gravity}\ }\textbf {\bibinfo {volume} {37}},\
  \bibinfo {pages} {205008} (\bibinfo {year} {2020})}\BibitemShut {NoStop}%
\bibitem [{\citenamefont {Ke}\ \emph {et~al.}(2022)\citenamefont {Ke},
  \citenamefont {Luo}, \citenamefont {Tan}, \citenamefont {Liu}, \citenamefont
  {Shao},\ and\ \citenamefont {Yang}}]{Ke_2022}%
  \BibitemOpen
  \bibfield  {author} {\bibinfo {author} {\bibfnamefont {J.}~\bibnamefont
  {Ke}}, \bibinfo {author} {\bibfnamefont {J.}~\bibnamefont {Luo}}, \bibinfo
  {author} {\bibfnamefont {Y.-J.}\ \bibnamefont {Tan}}, \bibinfo {author}
  {\bibfnamefont {Z.}~\bibnamefont {Liu}}, \bibinfo {author} {\bibfnamefont
  {C.-G.}\ \bibnamefont {Shao}},\ and\ \bibinfo {author} {\bibfnamefont
  {S.-Q.}\ \bibnamefont {Yang}},\ }\bibfield  {title} {\bibinfo {title}
  {Non-isothermal squeeze film damping in the test of gravitational
  inverse-square law},\ }\href {https://doi.org/10.1088/1361-6382/ac6479}
  {\bibfield  {journal} {\bibinfo  {journal} {Classical Quantum Gravity}\
  }\textbf {\bibinfo {volume} {39}},\ \bibinfo {pages} {115004} (\bibinfo
  {year} {2022})}\BibitemShut {NoStop}%
\bibitem [{\citenamefont {Mao}\ \emph {et~al.}(2023)\citenamefont {Mao},
  \citenamefont {Tan}, \citenamefont {Liu}, \citenamefont {Yang}, \citenamefont
  {Luo}, \citenamefont {Shao},\ and\ \citenamefont {Zhou}}]{Mao_2023}%
  \BibitemOpen
  \bibfield  {author} {\bibinfo {author} {\bibfnamefont {J.-J.}\ \bibnamefont
  {Mao}}, \bibinfo {author} {\bibfnamefont {Y.-J.}\ \bibnamefont {Tan}},
  \bibinfo {author} {\bibfnamefont {J.-P.}\ \bibnamefont {Liu}}, \bibinfo
  {author} {\bibfnamefont {S.-Q.}\ \bibnamefont {Yang}}, \bibinfo {author}
  {\bibfnamefont {J.}~\bibnamefont {Luo}}, \bibinfo {author} {\bibfnamefont
  {C.-G.}\ \bibnamefont {Shao}},\ and\ \bibinfo {author} {\bibfnamefont
  {Z.-B.}\ \bibnamefont {Zhou}},\ }\bibfield  {title} {\bibinfo {title}
  {Residual gas damping noise in constrained volume in space-borne
  gravitational wave detection},\ }\href
  {https://doi.org/10.1088/1361-6382/acc167} {\bibfield  {journal} {\bibinfo
  {journal} {Classical Quantum Gravity}\ }\textbf {\bibinfo {volume} {40}},\
  \bibinfo {pages} {075015} (\bibinfo {year} {2023})}\BibitemShut {NoStop}%
\bibitem [{\citenamefont {Zhao}\ \emph {et~al.}(2023)\citenamefont {Zhao},
  \citenamefont {Li}, \citenamefont {Liu}, \citenamefont {Shao}, \citenamefont
  {Tan}, \citenamefont {Yin},\ and\ \citenamefont
  {Zhou}}]{PhysRevApplied.19.044005}%
  \BibitemOpen
  \bibfield  {author} {\bibinfo {author} {\bibfnamefont {Y.-J.}\ \bibnamefont
  {Zhao}}, \bibinfo {author} {\bibfnamefont {G.-L.}\ \bibnamefont {Li}},
  \bibinfo {author} {\bibfnamefont {L.}~\bibnamefont {Liu}}, \bibinfo {author}
  {\bibfnamefont {C.-G.}\ \bibnamefont {Shao}}, \bibinfo {author}
  {\bibfnamefont {D.-Y.}\ \bibnamefont {Tan}}, \bibinfo {author} {\bibfnamefont
  {H.}~\bibnamefont {Yin}},\ and\ \bibinfo {author} {\bibfnamefont {Z.-B.}\
  \bibnamefont {Zhou}},\ }\bibfield  {title} {\bibinfo {title} {Experimental
  verification of and physical interpretation for adsorption-dependent
  squeeze-film damping},\ }\href
  {https://doi.org/10.1103/PhysRevApplied.19.044005} {\bibfield  {journal}
  {\bibinfo  {journal} {Phys. Rev. Appl.}\ }\textbf {\bibinfo {volume} {19}},\
  \bibinfo {pages} {044005} (\bibinfo {year} {2023})}\BibitemShut {NoStop}%
\bibitem [{\citenamefont {Callen}\ and\ \citenamefont
  {Greene}(1952)}]{PhysRev.86.702}%
  \BibitemOpen
  \bibfield  {author} {\bibinfo {author} {\bibfnamefont {H.~B.}\ \bibnamefont
  {Callen}}\ and\ \bibinfo {author} {\bibfnamefont {R.~F.}\ \bibnamefont
  {Greene}},\ }\bibfield  {title} {\bibinfo {title} {On a theorem of
  irreversible thermodynamics},\ }\href
  {https://doi.org/10.1103/PhysRev.86.702} {\bibfield  {journal} {\bibinfo
  {journal} {Phys. Rev.}\ }\textbf {\bibinfo {volume} {86}},\ \bibinfo {pages}
  {702} (\bibinfo {year} {1952})}\BibitemShut {NoStop}%
\bibitem [{\citenamefont {Onsager}\ and\ \citenamefont
  {Machlup}(1953)}]{PhysRev.91.1505}%
  \BibitemOpen
  \bibfield  {author} {\bibinfo {author} {\bibfnamefont {L.}~\bibnamefont
  {Onsager}}\ and\ \bibinfo {author} {\bibfnamefont {S.}~\bibnamefont
  {Machlup}},\ }\bibfield  {title} {\bibinfo {title} {Fluctuations and
  irreversible processes},\ }\href {https://doi.org/10.1103/PhysRev.91.1505}
  {\bibfield  {journal} {\bibinfo  {journal} {Phys. Rev.}\ }\textbf {\bibinfo
  {volume} {91}},\ \bibinfo {pages} {1505} (\bibinfo {year}
  {1953})}\BibitemShut {NoStop}%
\bibitem [{\citenamefont {Kubo}(1966)}]{R_Kubo_1966}%
  \BibitemOpen
  \bibfield  {author} {\bibinfo {author} {\bibfnamefont {R.}~\bibnamefont
  {Kubo}},\ }\bibfield  {title} {\bibinfo {title} {The fluctuation-dissipation
  theorem},\ }\href {https://doi.org/10.1088/0034-4885/29/1/306} {\bibfield
  {journal} {\bibinfo  {journal} {Rep. Prog. Phys.}\ }\textbf {\bibinfo
  {volume} {29}},\ \bibinfo {pages} {255} (\bibinfo {year} {1966})}\BibitemShut
  {NoStop}%
\bibitem [{\citenamefont {Saulson}(1990)}]{PhysRevD.42.2437}%
  \BibitemOpen
  \bibfield  {author} {\bibinfo {author} {\bibfnamefont {P.~R.}\ \bibnamefont
  {Saulson}},\ }\bibfield  {title} {\bibinfo {title} {Thermal noise in
  mechanical experiments},\ }\href {https://doi.org/10.1103/PhysRevD.42.2437}
  {\bibfield  {journal} {\bibinfo  {journal} {Phys. Rev. D}\ }\textbf {\bibinfo
  {volume} {42}},\ \bibinfo {pages} {2437} (\bibinfo {year}
  {1990})}\BibitemShut {NoStop}%
\bibitem [{\citenamefont {Armano}\ \emph {et~al.}(2022)\citenamefont {Armano}
  \emph {et~al.}}]{PhysRevD.106.062001}%
  \BibitemOpen
  \bibfield  {author} {\bibinfo {author} {\bibfnamefont {M.}~\bibnamefont
  {Armano}} \emph {et~al.} (\bibinfo {collaboration} {LISA Pathfinder
  Collaboration}),\ }\bibfield  {title} {\bibinfo {title} {Transient
  acceleration events in {LISA Pathfinder} data: Properties and possible
  physical origin},\ }\href {https://doi.org/10.1103/PhysRevD.106.062001}
  {\bibfield  {journal} {\bibinfo  {journal} {Phys. Rev. D}\ }\textbf {\bibinfo
  {volume} {106}},\ \bibinfo {pages} {062001} (\bibinfo {year}
  {2022})}\BibitemShut {NoStop}%
\bibitem [{\citenamefont {Knudsen}(1909)}]{andp.19093330106}%
  \BibitemOpen
  \bibfield  {author} {\bibinfo {author} {\bibfnamefont {M.}~\bibnamefont
  {Knudsen}},\ }\bibfield  {title} {\bibinfo {title} {Die gesetze der
  molekularströmung und der inneren reibungsströmung der gase durch
  röhren},\ }\href {https://doi.org/https://doi.org/10.1002/andp.19093330106}
  {\bibfield  {journal} {\bibinfo  {journal} {Ann. Phys.}\ }\textbf {\bibinfo
  {volume} {333}},\ \bibinfo {pages} {75} (\bibinfo {year} {1909})}\BibitemShut
  {NoStop}%
\bibitem [{\citenamefont {Knudsen}(1967)}]{knudsen1967cosine}%
  \BibitemOpen
  \bibfield  {author} {\bibinfo {author} {\bibfnamefont {M.}~\bibnamefont
  {Knudsen}},\ }\href@noop {} {\emph {\bibinfo {title} {The Cosine Law in the
  Kinetic Theory of Gases}}}\ (\bibinfo  {publisher} {National Aeronautics and
  Space Administration},\ \bibinfo {year} {1967})\BibitemShut {NoStop}%
\bibitem [{\citenamefont {Knudsen}(1911)}]{knudsen1911molekularstromung}%
  \BibitemOpen
  \bibfield  {author} {\bibinfo {author} {\bibfnamefont {M.}~\bibnamefont
  {Knudsen}},\ }\bibfield  {title} {\bibinfo {title} {Molekularstr{\"o}mung des
  wasserstoffs durch r{\"o}hren und das hitzdrahtmanometer},\ }\href@noop {}
  {\bibfield  {journal} {\bibinfo  {journal} {Ann. Phys.}\ }\textbf {\bibinfo
  {volume} {340}},\ \bibinfo {pages} {389} (\bibinfo {year}
  {1911})}\BibitemShut {NoStop}%
\bibitem [{\citenamefont {Knudsen}(1917)}]{knudsen1917vaporisation}%
  \BibitemOpen
  \bibfield  {author} {\bibinfo {author} {\bibfnamefont {M.}~\bibnamefont
  {Knudsen}},\ }\bibfield  {title} {\bibinfo {title} {The vaporisation of
  crystal surfaces},\ }\href@noop {} {\bibfield  {journal} {\bibinfo  {journal}
  {Ann. Phys.}\ }\textbf {\bibinfo {volume} {52}},\ \bibinfo {pages} {105}
  (\bibinfo {year} {1917})}\BibitemShut {NoStop}%
\bibitem [{\citenamefont {Rodrigues}\ \emph {et~al.}(2003)\citenamefont
  {Rodrigues}, \citenamefont {Foulon}, \citenamefont {Liorzou},\ and\
  \citenamefont {Touboul}}]{Manuel_Rodrigues_2003}%
  \BibitemOpen
  \bibfield  {author} {\bibinfo {author} {\bibfnamefont {M.}~\bibnamefont
  {Rodrigues}}, \bibinfo {author} {\bibfnamefont {B.}~\bibnamefont {Foulon}},
  \bibinfo {author} {\bibfnamefont {F.}~\bibnamefont {Liorzou}},\ and\ \bibinfo
  {author} {\bibfnamefont {P.}~\bibnamefont {Touboul}},\ }\bibfield  {title}
  {\bibinfo {title} {Flight experience on {CHAMP and GRACE} with
  ultra-sensitive accelerometers and return for {LISA}},\ }\href
  {https://doi.org/10.1088/0264-9381/20/10/332} {\bibfield  {journal} {\bibinfo
   {journal} {Classical Quantum Gravity}\ }\textbf {\bibinfo {volume} {20}},\
  \bibinfo {pages} {S291} (\bibinfo {year} {2003})}\BibitemShut {NoStop}%
\bibitem [{\citenamefont {Marque}\ \emph {et~al.}(2008)\citenamefont {Marque},
  \citenamefont {Christophe}, \citenamefont {Liorzou}, \citenamefont
  {Bodovill{\'e}}, \citenamefont {Foulon}, \citenamefont {Gu{\'e}rard},\ and\
  \citenamefont {Lebat}}]{marque2008ultra}%
  \BibitemOpen
  \bibfield  {author} {\bibinfo {author} {\bibfnamefont {J.}~\bibnamefont
  {Marque}}, \bibinfo {author} {\bibfnamefont {B.}~\bibnamefont {Christophe}},
  \bibinfo {author} {\bibfnamefont {F.}~\bibnamefont {Liorzou}}, \bibinfo
  {author} {\bibfnamefont {G.}~\bibnamefont {Bodovill{\'e}}}, \bibinfo {author}
  {\bibfnamefont {B.}~\bibnamefont {Foulon}}, \bibinfo {author} {\bibfnamefont
  {J.}~\bibnamefont {Gu{\'e}rard}},\ and\ \bibinfo {author} {\bibfnamefont
  {V.}~\bibnamefont {Lebat}},\ }\bibfield  {title} {\bibinfo {title} {The ultra
  sensitive accelerometers of the {ESA GOCE} mission},\ }in\ \href@noop {}
  {\emph {\bibinfo {booktitle} {59th International Astronautical Congress
  (IAC-08-B1. 3.7), pp. TP}}},\ Vol.\ \bibinfo {volume} {137}\ (\bibinfo {year}
  {2008})\ p.\ \bibinfo {pages} {2668}\BibitemShut {NoStop}%
\bibitem [{\citenamefont {Suijlen}\ \emph {et~al.}(2009)\citenamefont
  {Suijlen}, \citenamefont {Koning}, \citenamefont {{van Gils}},\ and\
  \citenamefont {Beijerinck}}]{SUIJLEN2009171}%
  \BibitemOpen
  \bibfield  {author} {\bibinfo {author} {\bibfnamefont {M.}~\bibnamefont
  {Suijlen}}, \bibinfo {author} {\bibfnamefont {J.}~\bibnamefont {Koning}},
  \bibinfo {author} {\bibfnamefont {M.}~\bibnamefont {{van Gils}}},\ and\
  \bibinfo {author} {\bibfnamefont {H.}~\bibnamefont {Beijerinck}},\ }\bibfield
   {title} {\bibinfo {title} {Squeeze film damping in the free molecular flow
  regime with full thermal accommodation},\ }\href
  {https://doi.org/https://doi.org/10.1016/j.sna.2009.03.025} {\bibfield
  {journal} {\bibinfo  {journal} {Sens. Actuators, A}\ }\textbf {\bibinfo
  {volume} {156}},\ \bibinfo {pages} {171} (\bibinfo {year}
  {2009})}\BibitemShut {NoStop}%
\bibitem [{\citenamefont {Dolleman}\ \emph {et~al.}(2021)\citenamefont
  {Dolleman}, \citenamefont {Chakraborty}, \citenamefont {Ladiges},
  \citenamefont {van~der Zant}, \citenamefont {Sader},\ and\ \citenamefont
  {Steeneken}}]{acs.nanolett.1c02237}%
  \BibitemOpen
  \bibfield  {author} {\bibinfo {author} {\bibfnamefont {R.~J.}\ \bibnamefont
  {Dolleman}}, \bibinfo {author} {\bibfnamefont {D.}~\bibnamefont
  {Chakraborty}}, \bibinfo {author} {\bibfnamefont {D.~R.}\ \bibnamefont
  {Ladiges}}, \bibinfo {author} {\bibfnamefont {H.~S.~J.}\ \bibnamefont
  {van~der Zant}}, \bibinfo {author} {\bibfnamefont {J.~E.}\ \bibnamefont
  {Sader}},\ and\ \bibinfo {author} {\bibfnamefont {P.~G.}\ \bibnamefont
  {Steeneken}},\ }\bibfield  {title} {\bibinfo {title} {Squeeze-film effect on
  atomically thin resonators in the high-pressure limit},\ }\href
  {https://doi.org/10.1021/acs.nanolett.1c02237} {\bibfield  {journal}
  {\bibinfo  {journal} {Nano Lett.}\ }\textbf {\bibinfo {volume} {21}},\
  \bibinfo {pages} {7617} (\bibinfo {year} {2021})}\BibitemShut {NoStop}%
\bibitem [{\citenamefont {Li}\ and\ \citenamefont
  {Hughes}(2000)}]{10.1117/12.395618}%
  \BibitemOpen
  \bibfield  {author} {\bibinfo {author} {\bibfnamefont {G.~X.}\ \bibnamefont
  {Li}}\ and\ \bibinfo {author} {\bibfnamefont {H.~G.}\ \bibnamefont
  {Hughes}},\ }\bibfield  {title} {\bibinfo {title} {Review of viscous damping
  in micromachined structures},\ }in\ \href {https://doi.org/10.1117/12.395618}
  {\emph {\bibinfo {booktitle} {Micromachined Devices and Components VI}}},\
  Vol.\ \bibinfo {volume} {4176}\ (\bibinfo  {publisher} {SPIE},\ \bibinfo
  {year} {2000})\ pp.\ \bibinfo {pages} {30 -- 46}\BibitemShut {NoStop}%
\bibitem [{\citenamefont {Andrews}\ \emph {et~al.}(1993)\citenamefont
  {Andrews}, \citenamefont {Harris},\ and\ \citenamefont
  {Turner}}]{ANDREWS199379}%
  \BibitemOpen
  \bibfield  {author} {\bibinfo {author} {\bibfnamefont {M.}~\bibnamefont
  {Andrews}}, \bibinfo {author} {\bibfnamefont {I.}~\bibnamefont {Harris}},\
  and\ \bibinfo {author} {\bibfnamefont {G.}~\bibnamefont {Turner}},\
  }\bibfield  {title} {\bibinfo {title} {A comparison of squeeze-film theory
  with measurements on a microstructure},\ }\href
  {https://doi.org/https://doi.org/10.1016/0924-4247(93)80144-6} {\bibfield
  {journal} {\bibinfo  {journal} {Sens. Actuators, A}\ }\textbf {\bibinfo
  {volume} {36}},\ \bibinfo {pages} {79} (\bibinfo {year} {1993})}\BibitemShut
  {NoStop}%
\bibitem [{\citenamefont {Andrews}\ and\ \citenamefont
  {Harris}(1995)}]{ANDREWS1995103}%
  \BibitemOpen
  \bibfield  {author} {\bibinfo {author} {\bibfnamefont {M.}~\bibnamefont
  {Andrews}}\ and\ \bibinfo {author} {\bibfnamefont {P.}~\bibnamefont
  {Harris}},\ }\bibfield  {title} {\bibinfo {title} {Damping and gas viscosity
  measurements using a microstructure},\ }\href
  {https://doi.org/https://doi.org/10.1016/0924-4247(95)01005-L} {\bibfield
  {journal} {\bibinfo  {journal} {Sens. Actuators, A}\ }\textbf {\bibinfo
  {volume} {49}},\ \bibinfo {pages} {103} (\bibinfo {year} {1995})}\BibitemShut
  {NoStop}%
\bibitem [{\citenamefont {Kadar}\ \emph {et~al.}(1995)\citenamefont {Kadar},
  \citenamefont {Kindt}, \citenamefont {Bossche},\ and\ \citenamefont
  {Mollinger}}]{721736}%
  \BibitemOpen
  \bibfield  {author} {\bibinfo {author} {\bibfnamefont {Z.}~\bibnamefont
  {Kadar}}, \bibinfo {author} {\bibfnamefont {W.}~\bibnamefont {Kindt}},
  \bibinfo {author} {\bibfnamefont {A.}~\bibnamefont {Bossche}},\ and\ \bibinfo
  {author} {\bibfnamefont {J.}~\bibnamefont {Mollinger}},\ }\bibfield  {title}
  {\bibinfo {title} {Calculation of the quality factor of torsional resonators
  in the low-pressure region},\ }in\ \href
  {https://doi.org/10.1109/SENSOR.1995.721736} {\emph {\bibinfo {booktitle}
  {Proceedings of the International Solid-State Sensors and Actuators
  Conference - TRANSDUCERS' 95}}},\ Vol.~\bibinfo {volume} {2}\ (\bibinfo
  {year} {1995})\ pp.\ \bibinfo {pages} {29--32}\BibitemShut {NoStop}%
\bibitem [{\citenamefont {Li}\ \emph {et~al.}(1999)\citenamefont {Li},
  \citenamefont {Wu}, \citenamefont {Zhu},\ and\ \citenamefont
  {Liu}}]{LI1999191}%
  \BibitemOpen
  \bibfield  {author} {\bibinfo {author} {\bibfnamefont {B.}~\bibnamefont
  {Li}}, \bibinfo {author} {\bibfnamefont {H.}~\bibnamefont {Wu}}, \bibinfo
  {author} {\bibfnamefont {C.}~\bibnamefont {Zhu}},\ and\ \bibinfo {author}
  {\bibfnamefont {J.}~\bibnamefont {Liu}},\ }\bibfield  {title} {\bibinfo
  {title} {The theoretical analysis on damping characteristics of resonant
  microbeam in vacuum},\ }\href
  {https://doi.org/https://doi.org/10.1016/S0924-4247(99)00072-2} {\bibfield
  {journal} {\bibinfo  {journal} {Sens. Actuators, A}\ }\textbf {\bibinfo
  {volume} {77}},\ \bibinfo {pages} {191} (\bibinfo {year} {1999})}\BibitemShut
  {NoStop}%
\bibitem [{\citenamefont {Zook}\ \emph {et~al.}(1992)\citenamefont {Zook},
  \citenamefont {Burns}, \citenamefont {Guckel}, \citenamefont {Sniegowski},
  \citenamefont {Engelstad},\ and\ \citenamefont {Feng}}]{ZOOK199251}%
  \BibitemOpen
  \bibfield  {author} {\bibinfo {author} {\bibfnamefont {J.}~\bibnamefont
  {Zook}}, \bibinfo {author} {\bibfnamefont {D.}~\bibnamefont {Burns}},
  \bibinfo {author} {\bibfnamefont {H.}~\bibnamefont {Guckel}}, \bibinfo
  {author} {\bibfnamefont {J.}~\bibnamefont {Sniegowski}}, \bibinfo {author}
  {\bibfnamefont {R.}~\bibnamefont {Engelstad}},\ and\ \bibinfo {author}
  {\bibfnamefont {Z.}~\bibnamefont {Feng}},\ }\bibfield  {title} {\bibinfo
  {title} {Characteristics of polysilicon resonant microbeams},\ }\href
  {https://doi.org/https://doi.org/10.1016/0924-4247(92)87007-4} {\bibfield
  {journal} {\bibinfo  {journal} {Sens. Actuators, A}\ }\textbf {\bibinfo
  {volume} {35}},\ \bibinfo {pages} {51} (\bibinfo {year} {1992})}\BibitemShut
  {NoStop}%
\bibitem [{\citenamefont {Bao}\ \emph {et~al.}(2002)\citenamefont {Bao},
  \citenamefont {Yang}, \citenamefont {Yin},\ and\ \citenamefont
  {Sun}}]{MinhangBao_2002}%
  \BibitemOpen
  \bibfield  {author} {\bibinfo {author} {\bibfnamefont {M.}~\bibnamefont
  {Bao}}, \bibinfo {author} {\bibfnamefont {H.}~\bibnamefont {Yang}}, \bibinfo
  {author} {\bibfnamefont {H.}~\bibnamefont {Yin}},\ and\ \bibinfo {author}
  {\bibfnamefont {Y.}~\bibnamefont {Sun}},\ }\bibfield  {title} {\bibinfo
  {title} {Energy transfer model for squeeze-film air damping in low vacuum},\
  }\href {https://doi.org/10.1088/0960-1317/12/3/322} {\bibfield  {journal}
  {\bibinfo  {journal} {J. Micromech. Microeng.}\ }\textbf {\bibinfo {volume}
  {12}},\ \bibinfo {pages} {341} (\bibinfo {year} {2002})}\BibitemShut
  {NoStop}%
\bibitem [{\citenamefont {Lu}\ \emph {et~al.}(2018)\citenamefont {Lu},
  \citenamefont {Li}, \citenamefont {Bao},\ and\ \citenamefont
  {Fang}}]{Lu_2018}%
  \BibitemOpen
  \bibfield  {author} {\bibinfo {author} {\bibfnamefont {C.}~\bibnamefont
  {Lu}}, \bibinfo {author} {\bibfnamefont {P.}~\bibnamefont {Li}}, \bibinfo
  {author} {\bibfnamefont {M.}~\bibnamefont {Bao}},\ and\ \bibinfo {author}
  {\bibfnamefont {Y.}~\bibnamefont {Fang}},\ }\bibfield  {title} {\bibinfo
  {title} {A generalized energy transfer model for squeeze-film air damping in
  the free molecular regime},\ }\href
  {https://doi.org/10.1088/1361-6439/aabdc0} {\bibfield  {journal} {\bibinfo
  {journal} {J. Micromech. Microeng.}\ }\textbf {\bibinfo {volume} {28}},\
  \bibinfo {pages} {085003} (\bibinfo {year} {2018})}\BibitemShut {NoStop}%
\bibitem [{\citenamefont {Yang}\ \emph {et~al.}(2009)\citenamefont {Yang},
  \citenamefont {Cheng}, \citenamefont {Dai}, \citenamefont {Li},\ and\
  \citenamefont {Wang}}]{5398187}%
  \BibitemOpen
  \bibfield  {author} {\bibinfo {author} {\bibfnamefont {H.}~\bibnamefont
  {Yang}}, \bibinfo {author} {\bibfnamefont {H.}~\bibnamefont {Cheng}},
  \bibinfo {author} {\bibfnamefont {B.}~\bibnamefont {Dai}}, \bibinfo {author}
  {\bibfnamefont {X.}~\bibnamefont {Li}},\ and\ \bibinfo {author}
  {\bibfnamefont {Y.}~\bibnamefont {Wang}},\ }\bibfield  {title} {\bibinfo
  {title} {A non-isothermal model for squeeze film damping of rarefied gas},\
  }in\ \href {https://doi.org/10.1109/ICSENS.2009.5398187} {\emph {\bibinfo
  {booktitle} {SENSORS, 2009 IEEE}}}\ (\bibinfo {year} {2009})\ pp.\ \bibinfo
  {pages} {213--216}\BibitemShut {NoStop}%
\bibitem [{\citenamefont {Moore}\ \emph {et~al.}(2003)\citenamefont {Moore},
  \citenamefont {Turner},\ and\ \citenamefont {Qiang}}]{MOORE20031897}%
  \BibitemOpen
  \bibfield  {author} {\bibinfo {author} {\bibfnamefont {P.}~\bibnamefont
  {Moore}}, \bibinfo {author} {\bibfnamefont {J.}~\bibnamefont {Turner}},\ and\
  \bibinfo {author} {\bibfnamefont {Z.}~\bibnamefont {Qiang}},\ }\bibfield
  {title} {\bibinfo {title} {{Champ} orbit determination and gravity field
  recovery},\ }\href {https://doi.org/10.1016/S0273-1177(03)00164-9} {\bibfield
   {journal} {\bibinfo  {journal} {Adv. Space Res.}\ }\textbf {\bibinfo
  {volume} {31}},\ \bibinfo {pages} {1897} (\bibinfo {year}
  {2003})}\BibitemShut {NoStop}%
\bibitem [{\citenamefont {Wu}\ \emph {et~al.}(2022)\citenamefont {Wu},
  \citenamefont {Xu}, \citenamefont {Zhao}, \citenamefont {Qiang},
  \citenamefont {Luo},\ and\ \citenamefont {Wu}}]{wu2022global}%
  \BibitemOpen
  \bibfield  {author} {\bibinfo {author} {\bibfnamefont {L.}~\bibnamefont
  {Wu}}, \bibinfo {author} {\bibfnamefont {P.}~\bibnamefont {Xu}}, \bibinfo
  {author} {\bibfnamefont {S.}~\bibnamefont {Zhao}}, \bibinfo {author}
  {\bibfnamefont {L.-E.}\ \bibnamefont {Qiang}}, \bibinfo {author}
  {\bibfnamefont {Z.}~\bibnamefont {Luo}},\ and\ \bibinfo {author}
  {\bibfnamefont {Y.}~\bibnamefont {Wu}},\ }\bibfield  {title} {\bibinfo
  {title} {Global gravity field model from {Taiji-1} observations},\ }\href
  {https://doi.org/10.1007/s12217-022-09998-5} {\bibfield  {journal} {\bibinfo
  {journal} {Microgravity Sci. Technol.}\ }\textbf {\bibinfo {volume} {34}},\
  \bibinfo {pages} {77} (\bibinfo {year} {2022})}\BibitemShut {NoStop}%
\bibitem [{\citenamefont {Steckelmacher}(1966)}]{STECKELMACHER1966561}%
  \BibitemOpen
  \bibfield  {author} {\bibinfo {author} {\bibfnamefont {W.}~\bibnamefont
  {Steckelmacher}},\ }\bibfield  {title} {\bibinfo {title} {A review of the
  molecular flow conductance for systems of tubes and components and the
  measurement of pumping speed},\ }\href
  {https://doi.org/https://doi.org/10.1016/0042-207X(66)91416-3} {\bibfield
  {journal} {\bibinfo  {journal} {Vacuum}\ }\textbf {\bibinfo {volume} {16}},\
  \bibinfo {pages} {561} (\bibinfo {year} {1966})}\BibitemShut {NoStop}%
\bibitem [{\citenamefont {Marx}\ \emph {et~al.}(2003)\citenamefont {Marx},
  \citenamefont {Sell},\ and\ \citenamefont {Lester}}]{10.1063/1.1675763}%
  \BibitemOpen
  \bibfield  {author} {\bibinfo {author} {\bibfnamefont {W.~F.}\ \bibnamefont
  {Marx}}, \bibinfo {author} {\bibfnamefont {N.~J.}\ \bibnamefont {Sell}},\
  and\ \bibinfo {author} {\bibfnamefont {J.~E.}\ \bibnamefont {Lester}},\
  }\bibfield  {title} {\bibinfo {title} {Angular distribution of reaction and
  vaporization products of sodium chloride single crystals},\ }\href
  {https://doi.org/10.1063/1.1675763} {\bibfield  {journal} {\bibinfo
  {journal} {J. Chem. Phys.}\ }\textbf {\bibinfo {volume} {55}},\ \bibinfo
  {pages} {5835} (\bibinfo {year} {2003})}\BibitemShut {NoStop}%
\bibitem [{\citenamefont {Xu}\ \emph {et~al.}(2020)\citenamefont {Xu},
  \citenamefont {Treadway}, \citenamefont {Murray}, \citenamefont {Minton},
  \citenamefont {Malaska}, \citenamefont {Cable},\ and\ \citenamefont
  {Hofmann}}]{10.1063/5.0011958}%
  \BibitemOpen
  \bibfield  {author} {\bibinfo {author} {\bibfnamefont {C.}~\bibnamefont
  {Xu}}, \bibinfo {author} {\bibfnamefont {C.~M.}\ \bibnamefont {Treadway}},
  \bibinfo {author} {\bibfnamefont {V.~J.}\ \bibnamefont {Murray}}, \bibinfo
  {author} {\bibfnamefont {T.~K.}\ \bibnamefont {Minton}}, \bibinfo {author}
  {\bibfnamefont {M.~J.}\ \bibnamefont {Malaska}}, \bibinfo {author}
  {\bibfnamefont {M.~L.}\ \bibnamefont {Cable}},\ and\ \bibinfo {author}
  {\bibfnamefont {A.~E.}\ \bibnamefont {Hofmann}},\ }\bibfield  {title}
  {\bibinfo {title} {Inelastic scattering dynamics of naphthalene and
  2-octanone on highly oriented pyrolytic graphite},\ }\href
  {https://doi.org/10.1063/5.0011958} {\bibfield  {journal} {\bibinfo
  {journal} {J. Chem. Phys.}\ }\textbf {\bibinfo {volume} {152}},\ \bibinfo
  {pages} {244709} (\bibinfo {year} {2020})}\BibitemShut {NoStop}%
\bibitem [{\citenamefont {L{\"u}th}(2001)}]{luth2001solid}%
  \BibitemOpen
  \bibfield  {author} {\bibinfo {author} {\bibfnamefont {H.}~\bibnamefont
  {L{\"u}th}},\ }\href@noop {} {\emph {\bibinfo {title} {Solid Surfaces,
  Interfaces and Thin Films}}},\ Vol.~\bibinfo {volume} {4}\ (\bibinfo
  {publisher} {Springer},\ \bibinfo {year} {2001})\BibitemShut {NoStop}%
\bibitem [{\citenamefont {Hurlbut}(1959)}]{hurlbut1959molecular}%
  \BibitemOpen
  \bibfield  {author} {\bibinfo {author} {\bibfnamefont {F.}~\bibnamefont
  {Hurlbut}},\ }\bibfield  {title} {\bibinfo {title} {Molecular scattering at
  the solid surface},\ }in\ \href@noop {} {\emph {\bibinfo {booktitle} {Recent
  Research in Molecular Beams}}}\ (\bibinfo  {publisher} {Academic Press},\
  \bibinfo {year} {1959})\ pp.\ \bibinfo {pages} {145--156}\BibitemShut
  {NoStop}%
\bibitem [{\citenamefont {Wenaas}(2003)}]{10.1063/1.1674619}%
  \BibitemOpen
  \bibfield  {author} {\bibinfo {author} {\bibfnamefont {E.~P.}\ \bibnamefont
  {Wenaas}},\ }\bibfield  {title} {\bibinfo {title} {Equilibrium cosine law and
  scattering symmetry at the gas–surface interface},\ }\href
  {https://doi.org/10.1063/1.1674619} {\bibfield  {journal} {\bibinfo
  {journal} {J. Chem. Phys.}\ }\textbf {\bibinfo {volume} {54}},\ \bibinfo
  {pages} {376} (\bibinfo {year} {2003})}\BibitemShut {NoStop}%
\bibitem [{\citenamefont {Feres}\ and\ \citenamefont
  {Yablonsky}(2004)}]{FERES20041541}%
  \BibitemOpen
  \bibfield  {author} {\bibinfo {author} {\bibfnamefont {R.}~\bibnamefont
  {Feres}}\ and\ \bibinfo {author} {\bibfnamefont {G.}~\bibnamefont
  {Yablonsky}},\ }\bibfield  {title} {\bibinfo {title} {{Knudsen's cosine} law
  and random billiards},\ }\href
  {https://doi.org/https://doi.org/10.1016/j.ces.2004.01.016} {\bibfield
  {journal} {\bibinfo  {journal} {Chem. Eng. Sci.}\ }\textbf {\bibinfo {volume}
  {59}},\ \bibinfo {pages} {1541} (\bibinfo {year} {2004})}\BibitemShut
  {NoStop}%
\bibitem [{\citenamefont {Celestini}\ and\ \citenamefont
  {Mortessagne}(2008)}]{PhysRevE.77.021202}%
  \BibitemOpen
  \bibfield  {author} {\bibinfo {author} {\bibfnamefont {F.}~\bibnamefont
  {Celestini}}\ and\ \bibinfo {author} {\bibfnamefont {F.}~\bibnamefont
  {Mortessagne}},\ }\bibfield  {title} {\bibinfo {title} {Cosine law at the
  atomic scale: Toward realistic simulations of {Knudsen} diffusion},\ }\href
  {https://doi.org/10.1103/PhysRevE.77.021202} {\bibfield  {journal} {\bibinfo
  {journal} {Phys. Rev. E}\ }\textbf {\bibinfo {volume} {77}},\ \bibinfo
  {pages} {021202} (\bibinfo {year} {2008})}\BibitemShut {NoStop}%
\bibitem [{\citenamefont {Imboden}\ and\ \citenamefont
  {Mohanty}(2014)}]{IMBODEN201489}%
  \BibitemOpen
  \bibfield  {author} {\bibinfo {author} {\bibfnamefont {M.}~\bibnamefont
  {Imboden}}\ and\ \bibinfo {author} {\bibfnamefont {P.}~\bibnamefont
  {Mohanty}},\ }\bibfield  {title} {\bibinfo {title} {Dissipation in
  nanoelectromechanical systems},\ }\href
  {https://doi.org/https://doi.org/10.1016/j.physrep.2013.09.003} {\bibfield
  {journal} {\bibinfo  {journal} {Phys. Rep.}\ }\textbf {\bibinfo {volume}
  {534}},\ \bibinfo {pages} {89} (\bibinfo {year} {2014})}\BibitemShut
  {NoStop}%
\bibitem [{\citenamefont {Bao}\ and\ \citenamefont {Yang}(2007)}]{BAO20073}%
  \BibitemOpen
  \bibfield  {author} {\bibinfo {author} {\bibfnamefont {M.}~\bibnamefont
  {Bao}}\ and\ \bibinfo {author} {\bibfnamefont {H.}~\bibnamefont {Yang}},\
  }\bibfield  {title} {\bibinfo {title} {Squeeze film air damping in {MEMS}},\
  }\href {https://doi.org/https://doi.org/10.1016/j.sna.2007.01.008} {\bibfield
   {journal} {\bibinfo  {journal} {Sens. Actuators, A}\ }\textbf {\bibinfo
  {volume} {136}},\ \bibinfo {pages} {3} (\bibinfo {year} {2007})}\BibitemShut
  {NoStop}%
\bibitem [{\citenamefont {Chandra}\ and\ \citenamefont
  {Keblinski}(2020)}]{10.1063/5.0018726}%
  \BibitemOpen
  \bibfield  {author} {\bibinfo {author} {\bibfnamefont {A.}~\bibnamefont
  {Chandra}}\ and\ \bibinfo {author} {\bibfnamefont {P.}~\bibnamefont
  {Keblinski}},\ }\bibfield  {title} {\bibinfo {title} {Investigating the
  validity of {Schrage} relationships for water using molecular dynamics
  simulations},\ }\href {https://doi.org/10.1063/5.0018726} {\bibfield
  {journal} {\bibinfo  {journal} {J. Chem. Phys.}\ }\textbf {\bibinfo {volume}
  {153}},\ \bibinfo {pages} {124505} (\bibinfo {year} {2020})}\BibitemShut
  {NoStop}%
\bibitem [{\citenamefont {Comsa}\ and\ \citenamefont
  {David}(1985)}]{COMSA1985145}%
  \BibitemOpen
  \bibfield  {author} {\bibinfo {author} {\bibfnamefont {G.}~\bibnamefont
  {Comsa}}\ and\ \bibinfo {author} {\bibfnamefont {R.}~\bibnamefont {David}},\
  }\bibfield  {title} {\bibinfo {title} {Dynamical parameters of desorbing
  molecules},\ }\href
  {https://doi.org/https://doi.org/10.1016/0167-5729(85)90009-3} {\bibfield
  {journal} {\bibinfo  {journal} {Surf. Sci. Rep.}\ }\textbf {\bibinfo {volume}
  {5}},\ \bibinfo {pages} {145} (\bibinfo {year} {1985})}\BibitemShut {NoStop}%
\bibitem [{\citenamefont {Lafferty}\ and\ \citenamefont
  {Rubin}(1999)}]{lafferty1999foundations}%
  \BibitemOpen
  \bibfield  {author} {\bibinfo {author} {\bibfnamefont {J.~M.}\ \bibnamefont
  {Lafferty}}\ and\ \bibinfo {author} {\bibfnamefont {L.~G.}\ \bibnamefont
  {Rubin}},\ }\href@noop {} {\emph {\bibinfo {title} {Foundations of Vacuum
  Science and Technology}}}\ (\bibinfo  {publisher} {American Institute of
  Physics},\ \bibinfo {year} {1999})\BibitemShut {NoStop}%
\bibitem [{\citenamefont {O'Hanlon}\ and\ \citenamefont
  {Gessert}(2003)}]{o2023user}%
  \BibitemOpen
  \bibfield  {author} {\bibinfo {author} {\bibfnamefont {J.~F.}\ \bibnamefont
  {O'Hanlon}}\ and\ \bibinfo {author} {\bibfnamefont {T.~A.}\ \bibnamefont
  {Gessert}},\ }\href@noop {} {\emph {\bibinfo {title} {A User's Guide to
  Vacuum Technology}}}\ (\bibinfo  {publisher} {John Wiley \& Sons},\ \bibinfo
  {year} {2003})\BibitemShut {NoStop}%
\bibitem [{\citenamefont {Pathria}(2016)}]{pathria2016statistical}%
  \BibitemOpen
  \bibfield  {author} {\bibinfo {author} {\bibfnamefont {R.~K.}\ \bibnamefont
  {Pathria}},\ }\href@noop {} {\emph {\bibinfo {title} {Statistical
  Mechanics}}}\ (\bibinfo  {publisher} {Elsevier},\ \bibinfo {year}
  {2016})\BibitemShut {NoStop}%
\bibitem [{\citenamefont {Touboul}(2001)}]{touboul2001space}%
  \BibitemOpen
  \bibfield  {author} {\bibinfo {author} {\bibfnamefont {P.}~\bibnamefont
  {Touboul}},\ }\bibfield  {title} {\bibinfo {title} {Space accelerometers:
  present status},\ }in\ \href@noop {} {\emph {\bibinfo {booktitle} {Gyros,
  Clocks, Interferometers...: Testing Relativistic Graviy in Space}}}\
  (\bibinfo  {publisher} {Springer},\ \bibinfo {year} {2001})\ pp.\ \bibinfo
  {pages} {273--291}\BibitemShut {NoStop}%
\bibitem [{\citenamefont {K{\"o}nig}(1751)}]{koenig1751universali}%
  \BibitemOpen
  \bibfield  {author} {\bibinfo {author} {\bibfnamefont {J.~S.}\ \bibnamefont
  {K{\"o}nig}},\ }\bibfield  {title} {\bibinfo {title} {De universali principio
  aequilibrii \& motus, in vi viva reperto, deque nexu inter vim vivam \&
  actionem, utriusque minimo, dissertatio},\ }\href@noop {} {\bibfield
  {journal} {\bibinfo  {journal} {Nova acta eruditorum}\ }\textbf {\bibinfo
  {volume} {3}},\ \bibinfo {pages} {125} (\bibinfo {year} {1751})}\BibitemShut
  {NoStop}%
\bibitem [{\citenamefont {Maxwell}(2003)}]{maxwell2003illustrations}%
  \BibitemOpen
  \bibfield  {author} {\bibinfo {author} {\bibfnamefont {J.~C.}\ \bibnamefont
  {Maxwell}},\ }\bibfield  {title} {\bibinfo {title} {Illustrations of the
  dynamical theory of gases},\ }in\ \href@noop {} {\emph {\bibinfo {booktitle}
  {The Kinetic Theory of Gases: an Anthology of Classic Papers with Historical
  Commentary}}}\ (\bibinfo  {publisher} {World Scientific},\ \bibinfo {year}
  {2003})\ pp.\ \bibinfo {pages} {148--171}\BibitemShut {NoStop}%
\bibitem [{\citenamefont {Amaro-Seoane}\ \emph {et~al.}(2017)\citenamefont
  {Amaro-Seoane} \emph {et~al.}}]{amaroseoane2017laser}%
  \BibitemOpen
  \bibfield  {author} {\bibinfo {author} {\bibfnamefont {P.}~\bibnamefont
  {Amaro-Seoane}} \emph {et~al.},\ }\href {https://arxiv.org/abs/1702.00786}
  {\bibinfo {title} {Laser interferometer space antenna}} (\bibinfo {year}
  {2017}),\ \Eprint {https://arxiv.org/abs/1702.00786} {arXiv:1702.00786}
  \BibitemShut {NoStop}%
\bibitem [{\citenamefont {Adelberger}\ \emph {et~al.}(2003)\citenamefont
  {Adelberger}, \citenamefont {Heckel},\ and\ \citenamefont
  {Nelson}}]{doi:10.1146/annurev.nucl.53.041002.110503}%
  \BibitemOpen
  \bibfield  {author} {\bibinfo {author} {\bibfnamefont {E.}~\bibnamefont
  {Adelberger}}, \bibinfo {author} {\bibfnamefont {B.}~\bibnamefont {Heckel}},\
  and\ \bibinfo {author} {\bibfnamefont {A.}~\bibnamefont {Nelson}},\
  }\bibfield  {title} {\bibinfo {title} {Tests of the gravitational
  inverse-square law},\ }\href
  {https://doi.org/10.1146/annurev.nucl.53.041002.110503} {\bibfield  {journal}
  {\bibinfo  {journal} {Annu. Rev. Nucl. Part. Sci.}\ }\textbf {\bibinfo
  {volume} {53}},\ \bibinfo {pages} {77} (\bibinfo {year} {2003})}\BibitemShut
  {NoStop}%
\bibitem [{\citenamefont {Adelberger}\ \emph {et~al.}(2009)\citenamefont
  {Adelberger}, \citenamefont {Gundlach}, \citenamefont {Heckel}, \citenamefont
  {Hoedl},\ and\ \citenamefont {Schlamminger}}]{ADELBERGER2009102}%
  \BibitemOpen
  \bibfield  {author} {\bibinfo {author} {\bibfnamefont {E.}~\bibnamefont
  {Adelberger}}, \bibinfo {author} {\bibfnamefont {J.}~\bibnamefont
  {Gundlach}}, \bibinfo {author} {\bibfnamefont {B.}~\bibnamefont {Heckel}},
  \bibinfo {author} {\bibfnamefont {S.}~\bibnamefont {Hoedl}},\ and\ \bibinfo
  {author} {\bibfnamefont {S.}~\bibnamefont {Schlamminger}},\ }\bibfield
  {title} {\bibinfo {title} {Torsion balance experiments: A low-energy frontier
  of particle physics},\ }\href
  {https://doi.org/https://doi.org/10.1016/j.ppnp.2008.08.002} {\bibfield
  {journal} {\bibinfo  {journal} {Prog. Part. Nucl. Phys.}\ }\textbf {\bibinfo
  {volume} {62}},\ \bibinfo {pages} {102} (\bibinfo {year} {2009})}\BibitemShut
  {NoStop}%
\bibitem [{\citenamefont {Luo}\ \emph {et~al.}(2020)\citenamefont {Luo} \emph
  {et~al.}}]{Luo_2020}%
  \BibitemOpen
  \bibfield  {author} {\bibinfo {author} {\bibfnamefont {J.}~\bibnamefont
  {Luo}} \emph {et~al.},\ }\bibfield  {title} {\bibinfo {title} {The first
  round result from the {TianQin-1} satellite},\ }\href
  {https://doi.org/10.1088/1361-6382/aba66a} {\bibfield  {journal} {\bibinfo
  {journal} {Classical Quantum Gravity}\ }\textbf {\bibinfo {volume} {37}},\
  \bibinfo {pages} {185013} (\bibinfo {year} {2020})}\BibitemShut {NoStop}%
\bibitem [{\citenamefont {Wu}\ \emph {et~al.}(2021)\citenamefont {Wu} \emph
  {et~al.}}]{taiji2021china}%
  \BibitemOpen
  \bibfield  {author} {\bibinfo {author} {\bibfnamefont {Y.-L.}\ \bibnamefont
  {Wu}} \emph {et~al.} (\bibinfo {collaboration} {The Taiji Scientific
  Collaboration}),\ }\bibfield  {title} {\bibinfo {title} {China’s first step
  towards probing the expanding universe and the nature of gravity using a
  space borne gravitational wave antenna},\ }\href
  {https://doi.org/10.1038/s42005-021-00529-z} {\bibfield  {journal} {\bibinfo
  {journal} {Commun. Phys.}\ }\textbf {\bibinfo {volume} {4}},\ \bibinfo
  {pages} {34} (\bibinfo {year} {2021})}\BibitemShut {NoStop}%
\bibitem [{\citenamefont {Fan}\ \emph {et~al.}(2019{\natexlab{a}})\citenamefont
  {Fan}, \citenamefont {Forsberg}, \citenamefont {Smith}, \citenamefont
  {Schr{\"o}der}, \citenamefont {Wagner}, \citenamefont {R{\"o}djeg{\aa}rd},
  \citenamefont {Fischer}, \citenamefont {{\"O}stling}, \citenamefont {Lemme},\
  and\ \citenamefont {Niklaus}}]{fan2019graphene}%
  \BibitemOpen
  \bibfield  {author} {\bibinfo {author} {\bibfnamefont {X.}~\bibnamefont
  {Fan}}, \bibinfo {author} {\bibfnamefont {F.}~\bibnamefont {Forsberg}},
  \bibinfo {author} {\bibfnamefont {A.~D.}\ \bibnamefont {Smith}}, \bibinfo
  {author} {\bibfnamefont {S.}~\bibnamefont {Schr{\"o}der}}, \bibinfo {author}
  {\bibfnamefont {S.}~\bibnamefont {Wagner}}, \bibinfo {author} {\bibfnamefont
  {H.}~\bibnamefont {R{\"o}djeg{\aa}rd}}, \bibinfo {author} {\bibfnamefont
  {A.~C.}\ \bibnamefont {Fischer}}, \bibinfo {author} {\bibfnamefont
  {M.}~\bibnamefont {{\"O}stling}}, \bibinfo {author} {\bibfnamefont {M.~C.}\
  \bibnamefont {Lemme}},\ and\ \bibinfo {author} {\bibfnamefont
  {F.}~\bibnamefont {Niklaus}},\ }\bibfield  {title} {\bibinfo {title}
  {Graphene ribbons with suspended masses as transducers in ultra-small
  nanoelectromechanical accelerometers},\ }\href
  {https://doi.org/10.1038/s41928-019-0287-1} {\bibfield  {journal} {\bibinfo
  {journal} {Nat. Electron.}\ }\textbf {\bibinfo {volume} {2}},\ \bibinfo
  {pages} {394} (\bibinfo {year} {2019}{\natexlab{a}})}\BibitemShut {NoStop}%
\bibitem [{\citenamefont {Fan}\ \emph {et~al.}(2019{\natexlab{b}})\citenamefont
  {Fan}, \citenamefont {Forsberg}, \citenamefont {Smith}, \citenamefont
  {Schröder}, \citenamefont {Wagner}, \citenamefont {Östling}, \citenamefont
  {Lemme},\ and\ \citenamefont {Niklaus}}]{acs.nanolett.9b01759}%
  \BibitemOpen
  \bibfield  {author} {\bibinfo {author} {\bibfnamefont {X.}~\bibnamefont
  {Fan}}, \bibinfo {author} {\bibfnamefont {F.}~\bibnamefont {Forsberg}},
  \bibinfo {author} {\bibfnamefont {A.~D.}\ \bibnamefont {Smith}}, \bibinfo
  {author} {\bibfnamefont {S.}~\bibnamefont {Schröder}}, \bibinfo {author}
  {\bibfnamefont {S.}~\bibnamefont {Wagner}}, \bibinfo {author} {\bibfnamefont
  {M.}~\bibnamefont {Östling}}, \bibinfo {author} {\bibfnamefont {M.~C.}\
  \bibnamefont {Lemme}},\ and\ \bibinfo {author} {\bibfnamefont
  {F.}~\bibnamefont {Niklaus}},\ }\bibfield  {title} {\bibinfo {title}
  {Suspended graphene membranes with attached silicon proof masses as
  piezoresistive nanoelectromechanical systems accelerometers},\ }\href
  {https://doi.org/10.1021/acs.nanolett.9b01759} {\bibfield  {journal}
  {\bibinfo  {journal} {Nano Lett.}\ }\textbf {\bibinfo {volume} {19}},\
  \bibinfo {pages} {6788} (\bibinfo {year} {2019}{\natexlab{b}})}\BibitemShut
  {NoStop}%
\bibitem [{\citenamefont {Bressi}\ \emph {et~al.}(2002)\citenamefont {Bressi},
  \citenamefont {Carugno}, \citenamefont {Onofrio},\ and\ \citenamefont
  {Ruoso}}]{PhysRevLett.88.041804}%
  \BibitemOpen
  \bibfield  {author} {\bibinfo {author} {\bibfnamefont {G.}~\bibnamefont
  {Bressi}}, \bibinfo {author} {\bibfnamefont {G.}~\bibnamefont {Carugno}},
  \bibinfo {author} {\bibfnamefont {R.}~\bibnamefont {Onofrio}},\ and\ \bibinfo
  {author} {\bibfnamefont {G.}~\bibnamefont {Ruoso}},\ }\bibfield  {title}
  {\bibinfo {title} {Measurement of the {Casimir} force between parallel
  metallic surfaces},\ }\href {https://doi.org/10.1103/PhysRevLett.88.041804}
  {\bibfield  {journal} {\bibinfo  {journal} {Phys. Rev. Lett.}\ }\textbf
  {\bibinfo {volume} {88}},\ \bibinfo {pages} {041804} (\bibinfo {year}
  {2002})}\BibitemShut {NoStop}%
\bibitem [{\citenamefont {Fong}\ \emph {et~al.}(2019)\citenamefont {Fong},
  \citenamefont {Li}, \citenamefont {Zhao}, \citenamefont {Yang}, \citenamefont
  {Wang},\ and\ \citenamefont {Zhang}}]{fong2019phonon}%
  \BibitemOpen
  \bibfield  {author} {\bibinfo {author} {\bibfnamefont {K.~Y.}\ \bibnamefont
  {Fong}}, \bibinfo {author} {\bibfnamefont {H.-K.}\ \bibnamefont {Li}},
  \bibinfo {author} {\bibfnamefont {R.}~\bibnamefont {Zhao}}, \bibinfo {author}
  {\bibfnamefont {S.}~\bibnamefont {Yang}}, \bibinfo {author} {\bibfnamefont
  {Y.}~\bibnamefont {Wang}},\ and\ \bibinfo {author} {\bibfnamefont
  {X.}~\bibnamefont {Zhang}},\ }\bibfield  {title} {\bibinfo {title} {Phonon
  heat transfer across a vacuum through quantum fluctuations},\ }\href
  {https://doi.org/10.1038/s41586-019-1800-4} {\bibfield  {journal} {\bibinfo
  {journal} {Nature}\ }\textbf {\bibinfo {volume} {576}},\ \bibinfo {pages}
  {243} (\bibinfo {year} {2019})}\BibitemShut {NoStop}%
\end{thebibliography}%

\end{document}